\documentclass[
 reprint,
superscriptaddress,
 amsmath,amssymb,
prb,
floatfix,
]{revtex4-2}

\usepackage{graphicx}
\usepackage{dcolumn}
\usepackage{bm}
\usepackage{color}
\usepackage{float}
\usepackage[colorlinks,urlcolor=blue,citecolor=blue,linkcolor=blue]{hyperref}

\begin{document}

\preprint{APS/123-QED}

\title{Wave packet dynamics in a non-Hermitian exciton-polariton system}

\author{Y.-M. Robin Hu}
\affiliation{%
ARC Centre of Excellence in Future Low-Energy Electronics Technologies and Department of Quantum Science and Technology, Research School of Physics, The Australian National University, Canberra, ACT 2601 Australia
}%
\author{Elena A. Ostrovskaya}%
\affiliation{%
ARC Centre of Excellence in Future Low-Energy Electronics Technologies and Department of Quantum Science and Technology, Research School of Physics, The Australian National University, Canberra, ACT 2601 Australia
}%
\author{Eliezer Estrecho}

\affiliation{%
ARC Centre of Excellence in Future Low-Energy Electronics Technologies and Department of Quantum Science and Technology, Research School of Physics, The Australian National University, Canberra, ACT 2601 Australia
}%

\date{\today}

\begin{abstract}
We theoretically investigate the dynamics of wave packets in a generic, non-Hermitian, optically anisotropic exciton-polariton system that exhibits degeneracies of its complex-valued eigenenergies in the form of pairs of exceptional points in momentum space. We observe the self-acceleration and reshaping of the wave packets governed by their eigenenergies. We further find that the exciton-polariton wave packets tend to self-organize into the eigenstate with the smaller decay rate, then propagate towards the minima of the decay rates in momentum space, resulting in directional transport in real space. We also describe the formation of pseudospin topological defects on the imaginary Fermi arc, where the decay rates of the two eigenstate coincide in momentum space. These effects of non-Hermiticity on the dynamics of exciton polaritons can be observed experimentally in a microcavity with optically anisotropic cavity spacer or exciton-hosting materials.
\end{abstract}

\maketitle


\section{\label{sec:level1}Introduction}
Open dissipative systems described by non-Hermitian Hamiltonian operators exhibit a special type of spectral degeneracies called exceptional points \cite{ghatak2019,bergholz2021,el-ganainy2018,ozdemir2019,gao2015,su2021}, leading to the emergence of novel topological invariants \cite{ghatak2019,bergholz2021,su2021}, new topological states \cite{ghatak2019,bergholz2021,el-ganainy2018,kunst2018,hofmann2020,weidemann2020}, nontrivial lasing \cite{ozdemir2019,jin2018,peng2014},  non-reciprocal transmission \cite{ozdemir2019,jin2018} and unidirectional transport \cite{longhi2015,kawabata2021}. Microcavity exciton polaritons, created when the electron-hole pairs (excitons) are strongly coupled to photons in an optical microcavity \cite{su2021,kasprzak2006,carusotto2013,deng2010}, represent an accessible solid-state platform for studies of non-Hermitian physics due to their inherent open-dissipative character. Non-Hermitian spectral degeneracies, both in parameter and momentum space, as well as the associated topological invariant, have been observed in this system \cite{gao2015,su2021,liao2021}.

In a Hermitian system, the motion of a wave packet is described by a pair of semi-classical equations of motion, and its centre-of-mass motion in momentum space is governed by external forces \cite{bleu2018wp,culcer2005,leblanc2021}. Recently, it was discovered that a wave packet in a system described by a non-Hermitian Dirac model can move in momentum space without the presence of an external force as a result of the growths and decay of its components corresponding to different eigenenergy branches \cite{solnyshkov2021}. The trajectories of the wave packets under this self-acceleration are polarization-dependent and the centre-of-mass momenta for certain initial conditions follow the gradient of the imaginary part of the eigenenergy. Similar effects were also described in the context of an one-dimensional non-Hermitian lattice  \cite{longhi2022}. In the absence of an out-of-plane magnetic field, the model of exciton polaritons in a planar microcavity has a band structure similar to the non-Hermitian Dirac model investigated in Ref. \cite{solnyshkov2021}, which is characterized by pairs of exceptional points connected by the so-called Fermi arcs \cite{su2021}. Therefore, we can expect similar wave packet dynamics to emerge in a non-Hermitian optically anisotropic exciton-polariton system.

In this work, we theoretically investigate the non-Hermitian wave packet dynamics in a microcavity exciton-polariton system. Apart from the motion in the absence of an external force, we also find that for some initial conditions, the wave packets tend to split into multiple components that propagate towards different directions. Moreover, these wave packets tend to self-organize into different eigenstates as they evolve and propagate towards the maxima of the imaginary part of the corresponding eigenenergy. Finally, we examine the exciton-polariton pseudospin textures resulting from the wave packet evolution and describe the emergent pseudospin anti-merons \cite{cilibrizzi2016,nagaosa2013,guo2020,zhang2021,borge2021,krol2021} on the imaginary Fermi arc in momentum space. The anti-merons are non-singular topological point defects that are characterized by half-integer topological invariant. They have been studied both in real space \cite{cilibrizzi2016,flayac2013,vishnevsky2013,krol2021} and in momentum space \cite{guo2020,mohanta2017} in a variety of physical systems. Their detection on the imaginary Fermi arc in an exciton-polariton system would signify a clear signature of the non-Hermitian wave packet dynamics.

This work is organized as follows. In Section \ref{sec: model}, we present the non-Hermitian exciton-polariton model considered in this work. In Section \ref{sec: NH SA}, we discuss the wave-packet self-acceleration and splitting in momentum space. In Section \ref{sec: real space WP}, we describe the asymptotic behaviour of the exciton-polariton wave packets in momentum space and the unidirectional propagation in real space. Finally, in Section \ref{sec: pseudospin}, we present our investigation on the dynamics of the exciton-polariton pseudospins, including the formation of the pseudospin defects on the imaginary Fermi arc without (Section \ref{sec: defect}) and with (Section \ref{sec: defect Delta}) the presence of an out-of-plane field, and the defects formation in real space (Section \ref{sec: defect r}). In Appendix \ref{sec: 2 band WP}, we compare the results for the exciton-polariton model presented here to those of the non-Hermitian Dirac model and the non-Hermitian Chern insulator.

\section{Theoretical Model}\label{sec: model}
In this work, we focus on a non-Hermitian exciton-polariton model exhibiting an effective non-Abelian gauge field. The general form of this 2$\times$2, two-band Hamiltonian in momentum space is $\mathbf{H}(\mathbf{k})={H}_0(\mathbf{k}){\mathbf{I}} + \overrightarrow{\mathbf{G}}(\mathbf{k})\cdot\overrightarrow{\sigma},$
where $\overrightarrow{\mathbf{G}}(\mathbf{k})=[G_x(\mathbf{k}),G_y(\mathbf{k}),G_z(\mathbf{k})]$ is the gauge field, $\overrightarrow{\sigma}$ is a vector of Pauli matrices, ${\mathbf{I}}$ is the $2\times 2$ identity matrix, and $\mathbf{k}=(k_x,k_y)$. The Hamiltonian is written in the basis of circular polarization of the cavity photon or the spin projection of the excitons on the $z$-direction normal to the plane of the exciton-hosting material embedded in the microcavity. This basis forms the psuedospin of the exciton polaritons. Dropping the $k$ dependence for brevity, the complex energy spectrum can be written simply as $E_\pm = H_0 \pm G$.

We model the exciton polaritons in an optical microcavity using the non-Hermitian Hamiltonian presented in Ref. \cite{su2021}, which has the components:
\begin{equation}
    \begin{split}\label{eq: Polariton H}
        H_0(\mathbf{k})&=E_0-i\gamma_0+\frac{\hbar^2 k^2}{2m}-i\chi{k}^4\\
        \overrightarrow{\mathbf{G}}(\mathbf{k})&=[\alpha-ia+(\beta-ib)(k_x^2-k_y^2),2k_x k_y(\beta-ib),\Delta].
    \end{split}
\end{equation}
The expression for the complex energy can be written as:
\begin{equation}\label{eq: Polariton E}
    \begin{split}
        E_\pm=&E_0-i\gamma_0+\frac{\hbar^2 k^2}{2m}-i\chi{k}^4\pm G\\
        G=&\bigg(\Big(\alpha-ia+(\beta-ib)(k_x+ik_y)^2\Big)\\&\times\Big(\alpha-ia+(\beta-ib)(k_x-ik_y)^2\Big)+\Delta^2\bigg)^{1/2}
    \end{split}
\end{equation}
where $G$ denotes the mean-subtracted eigenenergies. This model is a non-Hermitian generalization of the one describing the microcavity exciton polaritons in Ref. \cite{tercas2014}. Here, $\alpha$ denotes the microcavity anisotropy which leads to energy splitting between the linearly-polarized modes, $\beta$ describes the photonic spin-orbit coupling which splits the TE (transverse electric) and TM (transverse magnetic) modes \cite{tercas2014,su2021}. The parameters $a$, $b$ describe the photonic losses that are polarization-dependent due to microcavity anisotropy and the TE-TM splitting \cite{su2021}. The presence of these parameters results in the non-Hermitian nature of the Hamiltonian. The term $\Delta\sigma_z$ in the Hamiltonian describes the effects of Zeeman splitting induced by an out-of-plane magnetic field which can be introduced into the experimental setup \cite{klembt2018,su2021}. The term $\frac{\hbar^2 k^2}{2m}$ approximates the kinetic energies of the exciton polaritons, where $m$ is the effective polariton mass. The two terms $-i\gamma_0-i\chi{k}^4$ ensure that the imaginary part of the eigenenergies $\operatorname{Im}E_\pm$ are always negative. This is required since this model describes the effects of both gain (e.g., an optical pump) and loss (e.g., radiative decay) by an effective loss term which determines the linewidth of the exciton-polariton spectrum. Positive values of $\operatorname{Im}E_\pm$ would suggest a net gain, which is not physical for this system. 

\begin{figure}
    \centering
    \includegraphics[width=0.48\textwidth]{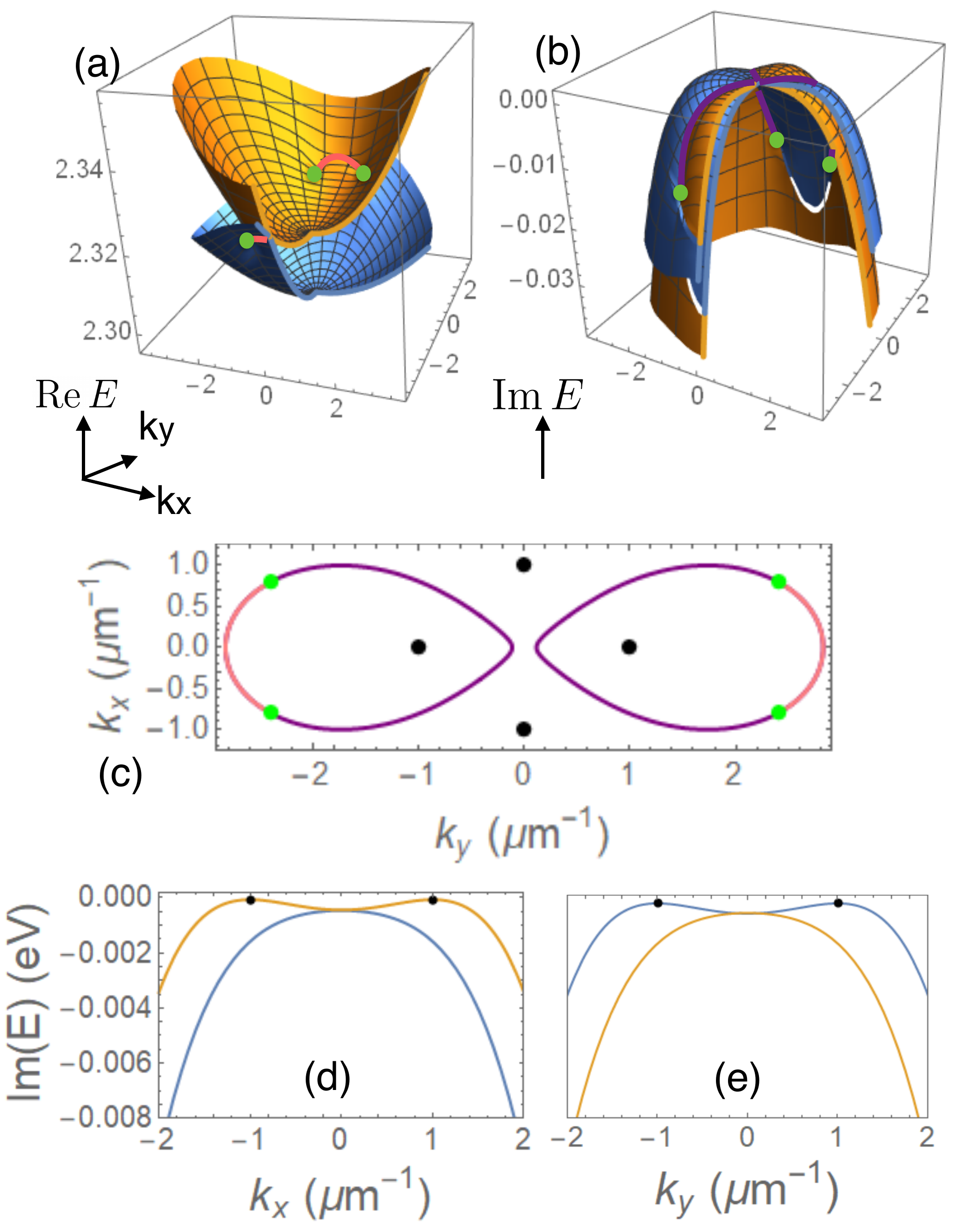}
    \caption{The (a) real part and (b) imaginary part of the eigenenergies of exciton-polariton model with cuts (denoted by the orange and blue lines) to show the degeneracies and the branch cuts. The green dots mark the exceptional points, the pink lines are the bulk Fermi arc where the two $\operatorname{Re}E$ surfaces touch and the purple lines are the imaginary Fermi arcs where the two $\operatorname{Im}E$ surfaces cross. (c): The simplified structure of the bulk Fermi arcs, the imaginary Fermi arcs and the exceptional points in momentum space. (d,e): The imaginary parts of the eigenenergies at $k_y=0$ and $k_x=0$, respectively. The black dots highlight the maxima of $\operatorname{Im}E_\pm$. The values of the parameters used here and the rest of the work are described in Appendix \ref{sec: para} unless specified otherwise.}
    \label{fig: 1}
\end{figure}

The eigenenergies $E_\pm$ are plotted in Fig. \ref{fig: 1}(a,b). The two exciton-polariton energy bands exhibit four exceptional points, each pair connected by the bulk Fermi arc where the real energy surfaces cross, and by the imaginary Fermi arc where the linewidth surfaces cross [see Fig. \ref{fig: 1}(a-c)] \cite{su2021, zhou2018}. A non-zero Zeeman splitting $\Delta$ would shrink the bulk Fermi arc, moving the exceptional points in a pair towards each other. A sufficiently strong Zeeman splitting, $|\Delta|>|(a\beta-b\alpha)/\beta|$, will destroy the exceptional points and open a gap. This shows that the exceptional points are more stable than the Dirac points (Hermitian spectral degeneracies) which annihilate for any non-zero value of $\Delta$.

The imaginary part of each eigenenergy branch has two local maxima. As shown in  Fig. \ref{fig: 1}(b) and highlighted in  Fig. \ref{fig: 1}(d,e), $\operatorname{Im}E_+$ (represented by the orange surfaces and the orange lines) has two maxima lying on the $k_y$-axis, while $\operatorname{Im}E_-$ (represented by the blue surfaces and the blue lines) has two maxima at the $k_x$-axis. These points play an important role in the dynamics of the exciton-polariton wave packets as discussed in the next Section.

\section{Results and Discussion}
Inspired by previous works \cite{solnyshkov2021,longhi2022}, we investigate the dynamics of Gaussian wave packets in a non-Hermitian exciton-polariton systems in the absence of an external potential. Since the Hamiltonian (\ref{eq: Polariton H}) contains $k$-components only, i.e., no external potential or spatial components, the time evolution of wave packets in real and momentum space and the resulting pseudospin textures can be calculated exactly (see Appendix \ref{sec: evolution} for details).

The initial exciton-polariton wave packet is Gaussian in momentum space and can be represented as a superposition of multiple $k$-components of the pseudospin eigenstates. The time evolution of the contributions from these components can be described as:
\begin{equation}
    \begin{split}
        |\psi^R_\pm(\mathbf{k},t)\rangle=e^{-i\operatorname{Re}E_\pm(\mathbf{k}) t} e^{\operatorname{Im}E_\pm(\mathbf{k}) t}|\psi^R_\pm(\mathbf{k})\rangle,
    \end{split}
\end{equation}
where the superscript $R$ denotes the right-eigenstates (see Appendix \ref{sec: eigenstates} for details). The real part of energies $\operatorname{Re}E_\pm$ describes the motion of the wave packet centre of mass in real space while the imaginary part $\operatorname{Im}E_\pm$ governs the growth and decay of the corresponding eigenstate. If $\operatorname{Im}E_\pm>0$, $\operatorname{Im}E_\pm$ corresponds to the growth rate of the corresponding eigenstate. Similarly, if $\operatorname{Im}E_\pm<0$, $-\operatorname{Im}E_\pm$ corresponds to the decay rate. The variation in the imaginary part, as shown below, is responsible for the peculiar effects described in this work.

\subsection{Self-Acceleration and Splitting of Exciton-Polariton Wave Packets}\label{sec: NH SA} 
\begin{figure}
    \centering
    \includegraphics[width=0.48\textwidth]{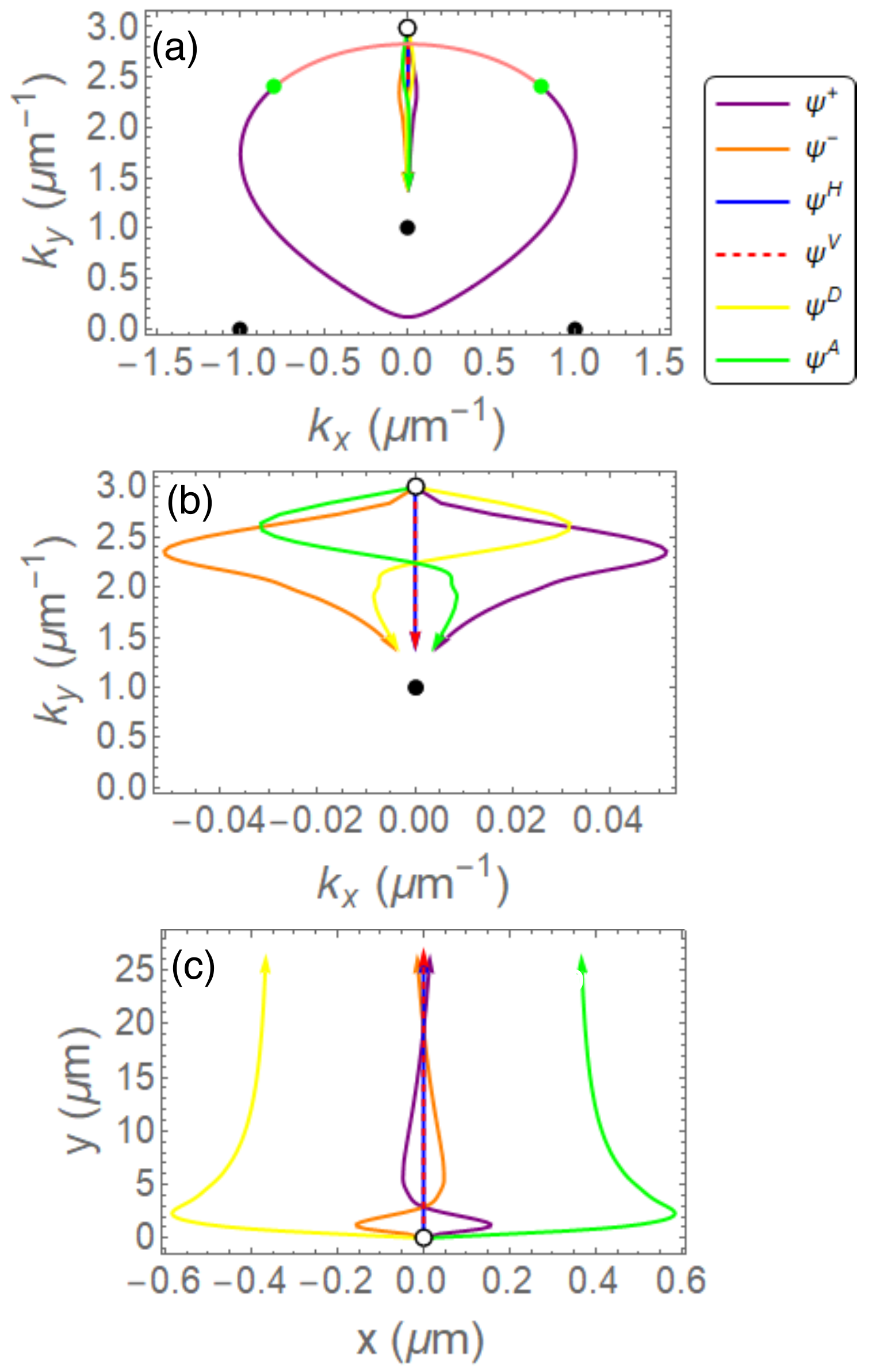}
    \caption{The polarization-dependent trajectories of the exciton-polariton wave-packet centre-of-mass in (a,b) the momentum space and (b) the real space, where (b) is the zoom-in of (a). Note that the trajectories of the horizontally-polarized and the vertically-polarized modes overlap. The white dots denote the initial wave packet centre of mass in both momentum and real space and the black dots denote the maxima of $\operatorname{Im}E_\pm$ in momentum space.}
    \label{fig: 2}
\end{figure}

Similar to the dynamics of the wave packets in a non-Hermitian Dirac model \cite{solnyshkov2021}, we observe the acceleration of the exciton-polariton wave packets in the absence of an external potential. Furthermore, the self-accelerating wave packet trajectories are sensitive to the initial polarization as shown by the centre-of-mass motion in Fig. \ref{fig: 2}(a-c). Note, however, that regardless of the initial polarization, self-accelerating wave packets tend to move towards the same point in momentum space or towards the same direction in real space. This acceleration arises from the gradient of the imaginary part of eigenenergy $\nabla \operatorname{Im} E_\pm$ [see Fig. \ref{fig: 1}(b,d,e)]. Namely, some $k$-components are decaying faster than others, resulting in an effective displacement of the wave packet center of mass towards $k$-components with larger imaginary part (or lesser decay rates).

\begin{figure}
    \centering
    \includegraphics[width=0.48\textwidth]{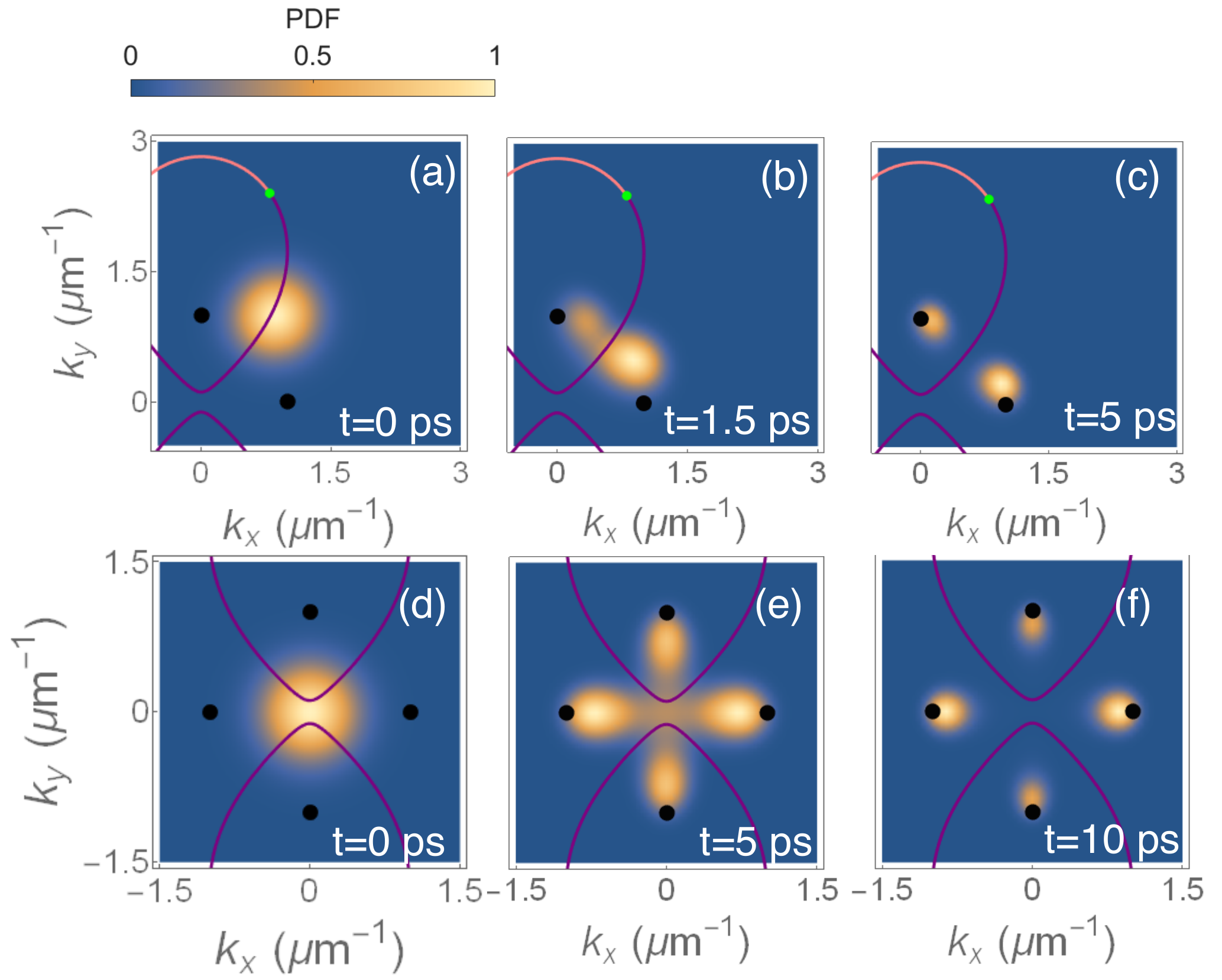}
    \caption{(a-c): The splitting of the exciton-polariton wave packet initially placed on the imaginary Fermi arc. (d-f): The splitting of the exciton-polariton wave packet initially placed at the origin. The black dots denote $\max(\operatorname{Im}E_\pm)$. The plots and the colorbar represent the evolution of the normalized probability density function (PDF). Note that this is not equal to the wave packet amplitude $|\psi|^2$, which decreases over time due to the continuous decay of the exciton polaritons.}
    \label{fig: 3}
\end{figure}

Surprisingly, when the exciton-polariton wave packet is initialized on the imaginary Fermi arc, it splits into two components that propagate away from each other as seen in Fig. \ref{fig: 3}(a-c). Furthermore, when the wave packet is initialized at the origin $\mathbf{k}=0$, while overlapping with the two imaginary Fermi arcs, it splits into four components. The four components accelerate away from each other along the $\pm k_x$ and $\pm k_y$ directions  [see Fig. \ref{fig: 3}(d-f)].

The splitting of the wave packets is due to the different imaginary parts (or decay rates) of the two pseudospin components. Since $\operatorname{Im}E_\pm$ corresponds to the growth/decay rates of the corresponding eigenstate, as time evolves, it is natural that the wave packet will self-organize towards the eigenstate with the larger growth rate (or the smaller decay rate) [see Fig. \ref{fig: 3}(a-f)]. In our model, $\operatorname{Im}E_+(\mathbf{k})$ is larger inside the region bounded by the Fermi arcs, while $\operatorname{Im}E_-(\mathbf{k})$ is larger outside of this region. Therefore, in the process of evolution, the components of the wave packets bounded by the Fermi arcs in momentum space will eventually self-organize into the upper eigenstate $|\psi_+^R(\mathbf{k})\rangle$, while the components outside the region tend to self-organize into the lower eigenstate $|\psi_-^R(\mathbf{k})\rangle$. Hence, when the wave packet splits in Fig. \ref{fig: 3}(a-c), the components belong to different eigenstates with the one inside the region bounded by the Fermi arcs belonging to the $|\psi_+^R(\mathbf{k})\rangle$, while the other to $|\psi_-^R(\mathbf{k})\rangle$. In a similar manner, for the case shown in Fig. \ref{fig: 3}(d-f), the two components inside (outside) the region bounded by the Fermi arcs belong to $|\psi_+^R(\mathbf{k})\rangle$ ($|\psi_-^R(\mathbf{k})\rangle$).

Note that since the difference in the imaginary parts of the two eigenenergies, $\Delta\operatorname{Im}E$, is larger for large $k$ [see Fig. \ref{fig: 1}], the components of the wave packets with large $k$ will self-organize into the two eigenstates faster than the components near $\mathbf{k}=0$. 

\begin{figure}
    \includegraphics[width=0.48\textwidth]{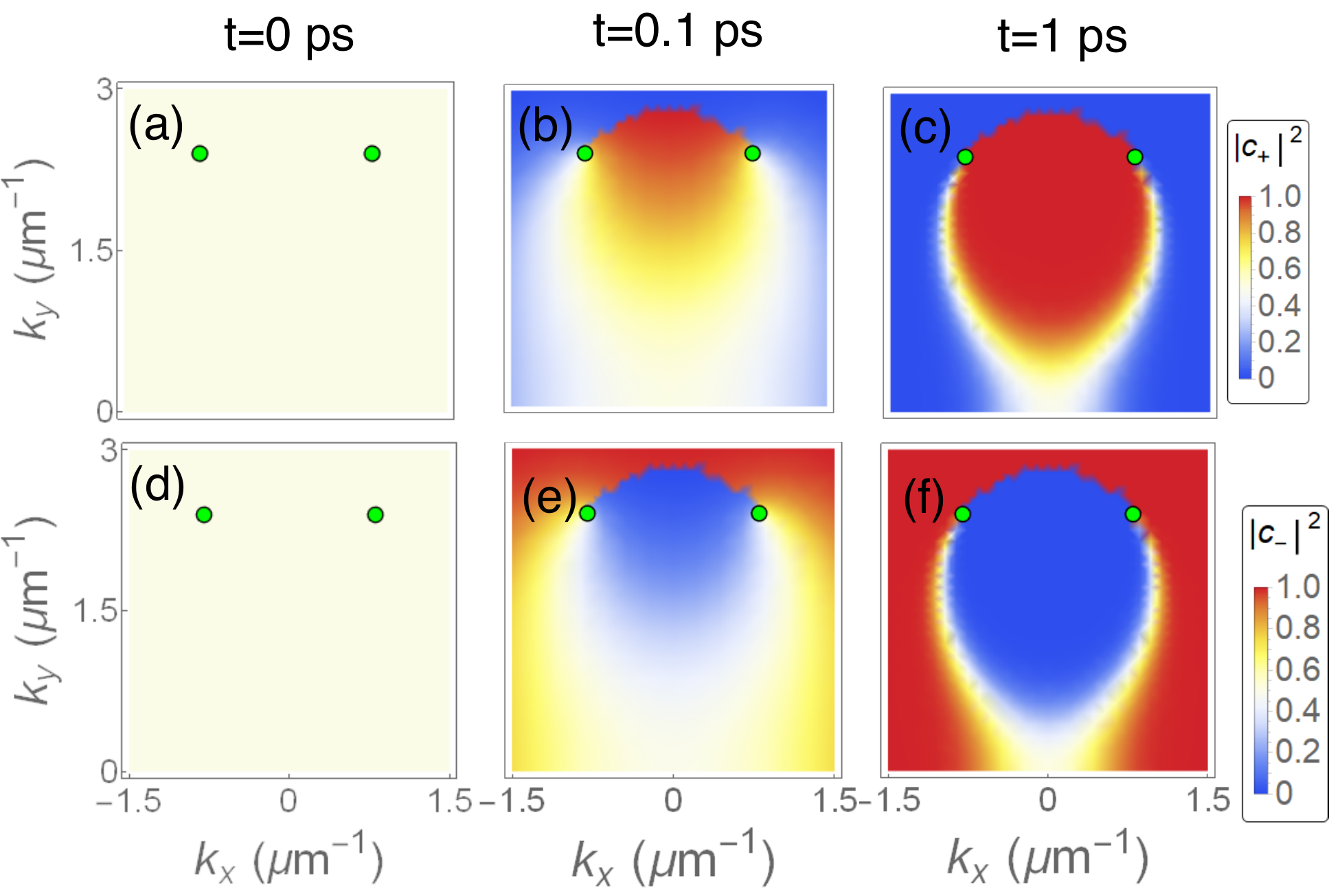}
    \caption{Evolution of the normalized (a-c) $|c_+|^2$ and (d-f) $|c_-|^2$ of the same wave packet, where the initial state of the wave packet is an equal superposition of the two eigenstate $|\psi\rangle=c_+|\psi_+^R\rangle+c_-|\psi_-^R\rangle$. (see Appendix \ref{sec: eigenstates} for details.) The green dots denote the exceptional points.}
    \label{fig: 4}
\end{figure}

\subsection{Directional Propagation of Exciton-Polariton Wave Packets}\label{sec: real space WP}
After the wave packet self-organizes into the eigenstates with the larger $\operatorname{Im}E$, its center-of-mass will then propagate towards the maxima of the corresponding $\operatorname{Im}E$ in momentum space. In real space, the wave packets will exhibit directional transport with a group velocity corresponding to the $\nabla_\mathbf{k}\operatorname{Re} E$ at $\mathbf{k}^*=\max(\operatorname{Im}E)$, similarly to the dynamics described in Ref. \cite{longhi2022}.

In our model, $\operatorname{Im}E_+$ has two maxima at $(0,\pm k^*_+)$ inside the region bounded by the Fermi arcs, while $\operatorname{Im}E_-$ has two maxima at $(\pm k^*_-,0)$ outside this region [see Fig. \ref{fig: 1}(c)]. The exciton-polariton wave packets that are inside the region bounded by the Fermi arcs will self-organize into the upper eigenstate and their center-of-mass momenta will reach the two fixed points at $\mathbf{k}^*=(0,\pm\sqrt{b/2\chi})$. On the other hand, the wave packets that are outside this region will self-organize into the lower eigenstate and their center-of-mass momenta will reach the other two fixed points at $\mathbf{k}^*=(\pm\sqrt{b/2\chi},0)$. In the real space, the wave packets in the upper eigenstate will asymptotically reach non-vanishing group velocities of 
\begin{equation}
    \langle\mathbf{v_+}\rangle=\bigg(0,\pm\Big(\frac{\hbar}{2m}-\beta/\hbar\Big)\sqrt{\frac{2b}{\chi}}\bigg)
\end{equation}
while the wave packets in the lower eigenstate will reach group velocities of
\begin{equation}
    \langle\mathbf{v_-}\rangle=\bigg(\pm\Big(\frac{\hbar}{2m}-\beta/\hbar\Big)\sqrt{\frac{2b}{\chi}},0\bigg).
\end{equation}
We note that these asymptotic group velocities of the exciton-polariton wave packets only depend on the mean linewidth $\chi$, the spin-orbit coupling parameter $\beta$, and the polarization-dependent loss $b$. 

\begin{figure}
    \centering
    \includegraphics[width=0.48\textwidth]{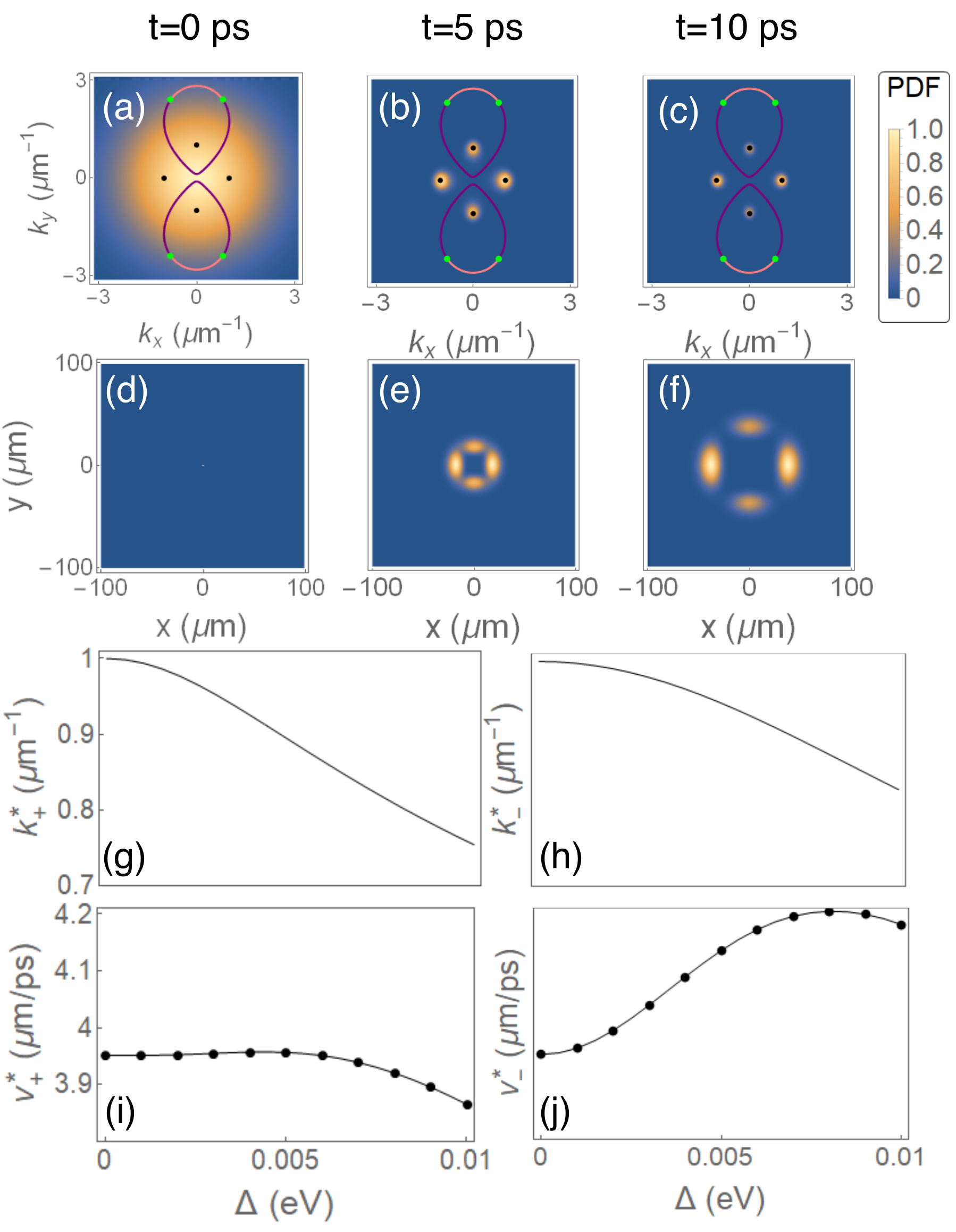}
    \caption{The time-evolution of a tight wave packet in (a-c) momentum space and (d-f) real space. (g) and (h) are the fixed points coordinates as a function $\Delta$ where $\mathbf{k}^*=(0,\pm k_+^*),(\pm k_-^*,0)$. (i) and (j) are the asymptotic group velocities where $\langle\mathbf{v}_+\rangle=(0,\pm v_+^*)$ and $\langle\mathbf{v}_-\rangle=(\pm v_-^*,0)$. The black lines in (i,j) represent the group velocities $\mathbf{v}=\nabla_\mathbf{k}\operatorname{Re}E_\pm$ calculated from $\mathbf{k}^*$ in (g,h), which show excellent agreement with the results from numerical simulation of wave packet dynamics (black dots).} 
    \label{fig: 5}
\end{figure}

Multiple extremum points $\mathbf{k}^*$ can be covered when an exciton-polariton wave packet is tightly focused in real space, since it is broad in momentum space. In this case, the wave packet will split into four parts with each part funnelling to the vicinity of $\mathbf{k}^*$ that it covered. In real space, the wave packet splits and spreads over time, eventually evolving into four broad wave packets in the two eigenstates. Each of the smaller wave packets exhibits directional transport with the corresponding non-vanishing asymptotic group velocities $\langle\mathbf{v_\pm}\rangle$ [see Fig. \ref{fig: 5}(a-f)].

In the presence of Zeeman splitting, i.e., $\Delta\neq 0$, the fixed points can be found numerically. Increasing $\Delta$ will move the fixed points towards $\mathbf{k}=0$ [see Fig. \ref{fig: 5}(g,h)]. Since the fixed points stay on the $k_x$ and $k_y$ axes, $\Delta$ will only change the magnitude of the final group velocities, but not the directions. The final group velocities depend on $\Delta$ non-monotonically, as seen in Fig. \ref{fig: 5}(i,j).

\begin{figure}[h]
    \centering
    \includegraphics[width=0.48\textwidth]{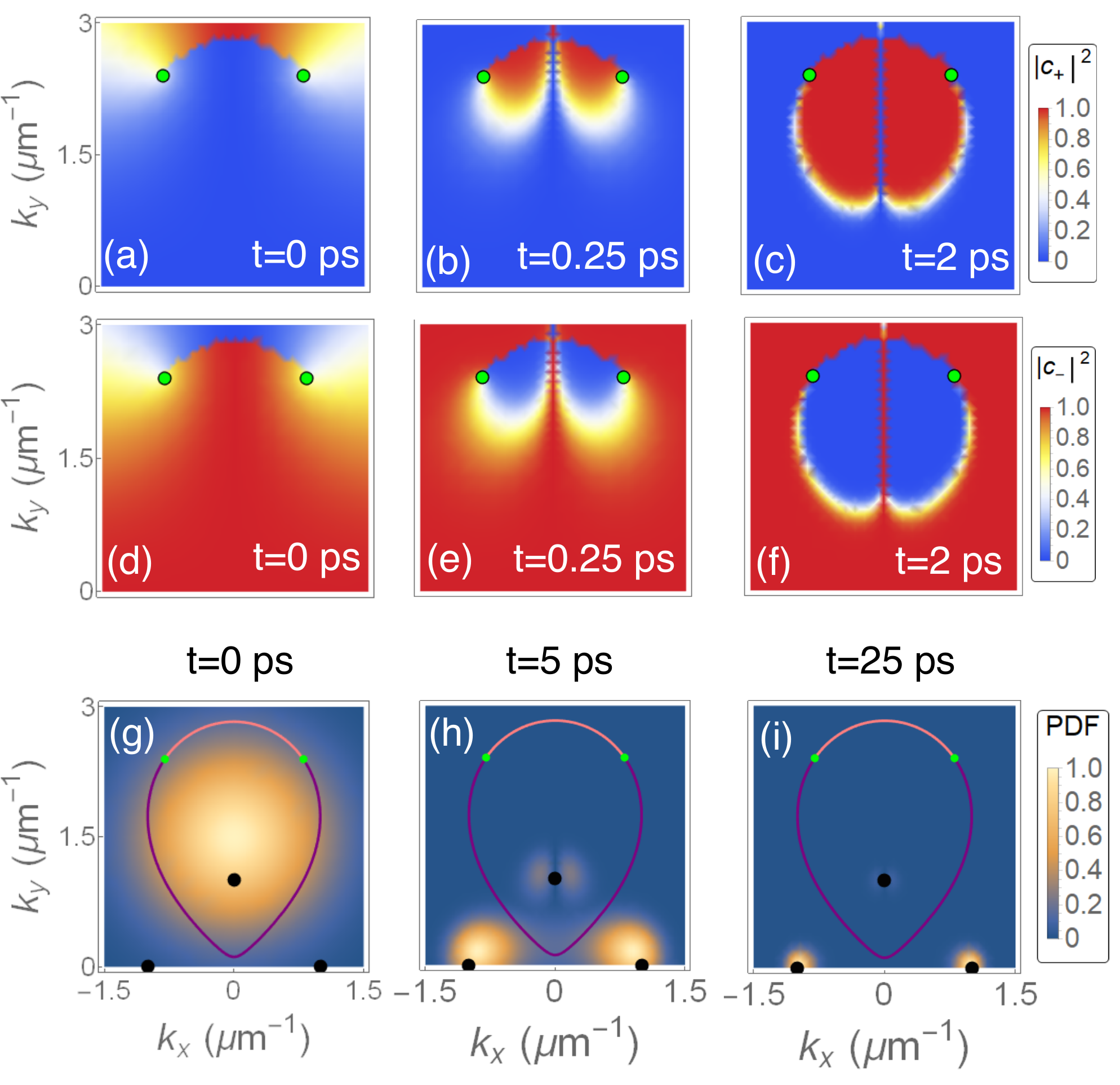}
    \caption{Time evolution of the normalized $|c_+|^2$ (a-c) and $|c_-|^2$ (d-f) when the initial wave packet is vertically-polarized. Also showing the time evolution of the wave packet (g-i) with black dots denoting $\mathbf{k}^*=\max(\operatorname{Im}E_\pm)$ and the green dots denote the exceptional points.}
    \label{fig: 6}
\end{figure}

It is worth mentioning that the above results are derived under the assumption that the exciton-polariton wave packet was initially in a superposition of the two eigenstates, $|\psi\rangle=c_+|\psi_+^R\rangle+c_-|\psi_-^R\rangle$, and have nonzero $|c_\pm|$ at all of $\mathbf{k}^*$. In contrast, if the initial state was in one of the eigenstate $|\psi^R_j\rangle$, where $j=+,-$, the exciton-polariton wave packet will only propagate towards the corresponding $\max(\operatorname{Im}E_j)$. Furthermore, if the components of the eigenstate $|c_j|^2$ in the initial state is zero at the corresponding $\max(\operatorname{Im}E_j)$, the wave packets will not funnel to $\max(\operatorname{Im}E_j)$ asymptotically. For example, when $\Delta=0$, the upper eigenstate at the fixed points $\mathbf{k}^*=(0,\pm\sqrt{b/2\chi})$ inside the region bounded by the Fermi arcs is horizontally polarized. Therefore, if the initial state is vertically polarized, which has no horizontal component, the exciton polaritons at $\mathbf{k}^*=(0,\pm\sqrt{b/2\chi})$ cannot self-organize into $|\psi_+^R\rangle$ as shown in Fig. \ref{fig: 6}(a-f). Consequently, the wave packet will never funnel to those fixed points [see Fig. \ref{fig: 6}(g-i)]. Instead, the wave packet will funnel to the other two fixed points at $\mathbf{k}^*=(\pm\sqrt{b/2\chi},0)$ and exhibit directional transport in real space at the  constant velocities $\langle\mathbf{v_-}\rangle$.

\subsection{Dynamics of Exciton-Polariton Pseudospin Textures}\label{sec: pseudospin}
The self-acceleration of the exciton-polariton wave packets is accompanied by a non-trivial pseudospin dynamics. The exciton-polariton pseudospin can be characterized by the Stokes vector $\mathbf{S}=[S_x,S_y,S_z]$ calculated using the polarization components as described in Appendix \ref{sec: S}, and is analogous to the electronic spins in a solid state system \cite{flayac2013,cilibrizzi2016,nagaosa2013,guo2020,zhang2021,stefano2016}. We note that the definitions of the in-plane Stokes vectors $[S_x,S_y]$ can be arbitrary, and the results in this Section are derived using the convention specified in Appendix \ref{sec: S}.

\begin{figure}
    \centering
    \includegraphics[width=0.48\textwidth]{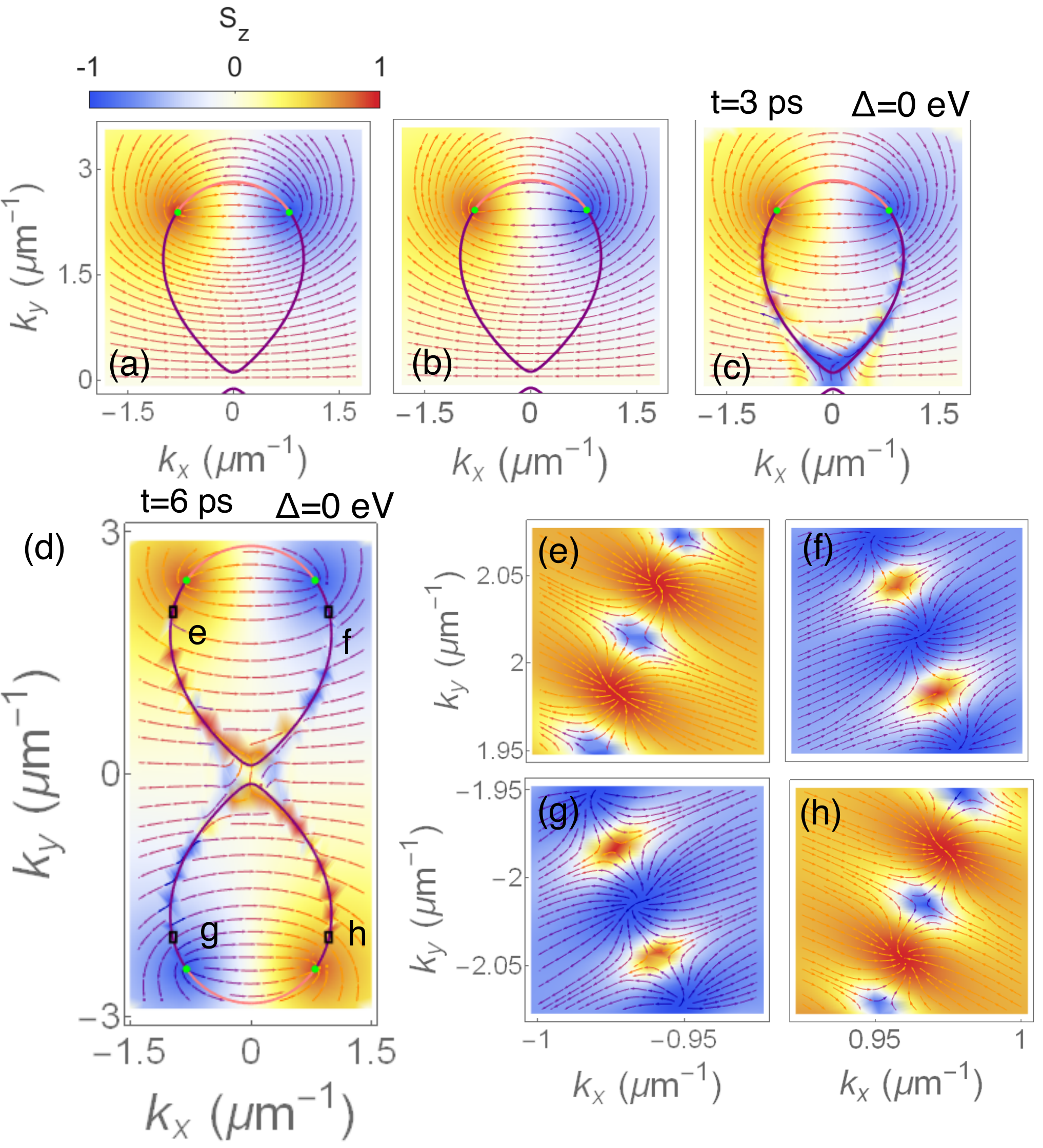}
    \caption{The pseudospin textures of (a) the upper eigenstate, (b) the lower eigenstate and (c) an exciton-polariton wave packet at $t=3$ ps with $\Delta=0$. (d) The pseudospin textures of the exciton-polariton wave packet at $t=6$ ps. (e-h): The zoom-in near the pseudospin defects on the imaginary Fermi arcs.}
    \label{fig: 7}
\end{figure}

As described above, the exciton-polariton wave packets tend to self-organize into different eigenstates in different regions in momentum space. Since these regions are separated by the Fermi arcs, it is natural to investigate what will happen on these Fermi arcs. We find that, despite the pseudospin textures of each eigenstate exhibiting discontinuous jumps on the bulk Fermi arcs, the pseudospin textures of one eigenstate can smoothly transition to the pseudospin textures of the other eigenstate across these arcs [see Fig. \ref{fig: 7}(a-c)]. On the other hand, the pseudospin textures of the two eigenstate are both continuous at the imaginary Fermi arcs, but they point in the opposite directions on these arcs [see Fig. \ref{fig: 7}(a-c)]. Furthermore, as the exciton-polariton pseudospins self-organize into different eigenstates, we observe the emergence of pseudospin defects resulting from the frustration. 

Note that the evolution of the pseudospin textures and the emergence of the defects depend only on the initial polarization of the wave packet, but not the initial wave packet widths, center-of-mass momenta, or the center-of-mass positions. Therefore, the pseudospin textures can evolve independently to the probability density function of the wave packet. Therefore, the defects can be generated even in the regions not covered by the bulk of the wave packet in momentum space.

\subsubsection{Pseudospin Defects}\label{sec: defect}
The pseudospin defects on the imaginary Fermi arcs are non-singular point defects similar to skyrmions \cite{everschor-sitte2018,borge2021,nagaosa2013,kovalev2018,flayac2013,cilibrizzi2016,guo2020,zhang2021,tretiakov2007,stefano2016,mohanta2017}. As the wave packet evolves, these defects first emerge at the segments of the imaginary Fermi arcs with large $k$, where the difference between the decay rates of the two bands, $\operatorname{Im}(\Delta E)$, is the largest. Then, more defects are generated at segments with smaller $k$ on the imaginary Fermi arc where $\operatorname{Im}(\Delta E)$ is smaller. The defects then move towards the exceptional points along the imaginary Fermi arc. This results in the defects becoming more densely packed as more defects are generated.

Topological point defects can be characterized by a winding number called the skyrmion number, $N_{sk}$, which counts how many times the pseudospin wraps around the unit sphere \cite{cilibrizzi2016,nagaosa2013,guo2020}:
\begin{equation}\label{eq: skyrmion}
\begin{split}
    N_{sk}&=\frac{1}{4\pi}\iint\mathbf{S}\cdot\Big(\partial_{x}\mathbf{S}\times\partial_{y}\mathbf{S}\Big)d^2\mathbf{r}.
\end{split}
\end{equation}
The skyrmion number can also be expressed in terms of two other topological invariants, the vorticity $w$ and the polarity $p$, as $N_{sk}=wp/2$. The vorticity measures the in-plane winding of the Stoke vector $\mathbf{S}$, where the angle of the in-plane Stoke vectors $[S_x,S_y]$ changes by $2\pi w$ along a closed curve encircling the core counter-clockwise \cite{borge2021,guo2020,krol2021,hertel2006}. For example, a spin vortex has a vorticity of $w=+1$, while an anti-vortex has a vorticity of $w=-1$ \cite{hertel2006}. On the other hand, the polarity measures the out-of-plane winding of $\mathbf{S}$ and quantifies the continuous change of $S_z$ from the centre to the edge of the defect \cite{borge2021,krol2021,guo2020,tretiakov2007,hertel2006,kovalev2018}. The Stoke vectors $\mathbf{S}$ in a skyrmion exhibits both the in-plane winding as well as the out-of-plane winding. The in-plane Stoke vectors of a skyrmion are similar to the ones of a vortex, resulting in a vorticity of $w=+1$. Furthermore, a core-up (core-down) skyrmion has an out-of-plane Stoke vector $S_z=+1$ ($S_z=-1$) at the centre, which continuously transforms into $S_z=-1$ ($S_z=+1$) at the edge, giving it a polarity of $p=+2$ ($p=-2$). \cite{everschor-sitte2018,borge2021,nagaosa2013,kovalev2018,flayac2013,cilibrizzi2016,guo2020,zhang2021,tretiakov2007,stefano2016}. Consequently, a core-up (core-down) skyrmion has a skyrmion number of $N_{sk}=+1$ ($N_{sk}=-1$). Similarly, an anti-skyrmion has in-plane Stoke vectors $[S_x,S_y]$ similar to the ones in an anti-vortex and a core-up (core-down) anti-skyrmion has a polarity of $p=+2$ ($p=-2$), giving it a skyrmion number of $N_{sk}=-1$ ($N_{sk}=+1$) \cite{guo2020}.

We observe core-up and core-down defects where the pseudospins lie on the $xy$-plane, that is $S_z=0$, on the edge of these defects [See Fig. \ref{fig: 7}(d-h)]. These defects have vorticities of $w=-1$ and the $z$-components of their pseudospins only perform a $\pi/2$-rotation instead of a $\pi$-rotation \cite{cilibrizzi2016}. Consequently, these core-up (core-down) defects have polarities of $p=+1$ ($p=-1$), giving them skyrmion numbers of $N_{sk}=-1/2$ ($N_{sk}=+1/2$). We therefore conclude that they are core-up and core-down anti-merons, i.e., anti-skyrmions with half-integer winding numbers. In the vicinity of the exceptional points in the upper-half plane ($k_y\geq0$), core-up and core-down anti-merons form on the right-hand side ($k_x\geq0$) and left-hand side ($k_x\leq0$) of the exceptional points pair, respectively. On the other hand, in the vicinity of the exceptional points in the lower-half plane ($k_y\leq0$), core-up and core-down anti-merons form on the left-hand side ($k_x\leq0$) and right-hand side ($k_x\geq0$) of the exceptional points pair, respectively. We also observe pseudospin vortices with core polarization $S_z=+1$ ($S_z=-1$) forming between the core-down (core-up) anti-merons on the imaginary Fermi arcs [see Fig. \ref{fig: 7}(d-h)]. The in-plane pseudospins $[S_x,S_y]$ wind around the vortex cores, giving them vorticities of $w=+1$. However, the pseudospin $S_z=\pm1$ at their cores does not transform into $S_z=\mp1$ or $S_z=0$ on the edge. Consequently, they do not have $p=\pm2$ like a skyrmion or $p=\pm1$ like a meron, and therefore do not have integer-valued or half-integer-valued skyrmion numbers.

\subsubsection{Pseudospin Defects with Zeeman Splitting}\label{sec: defect Delta}
\begin{figure}[h]
    \centering
    \includegraphics[width=0.48\textwidth]{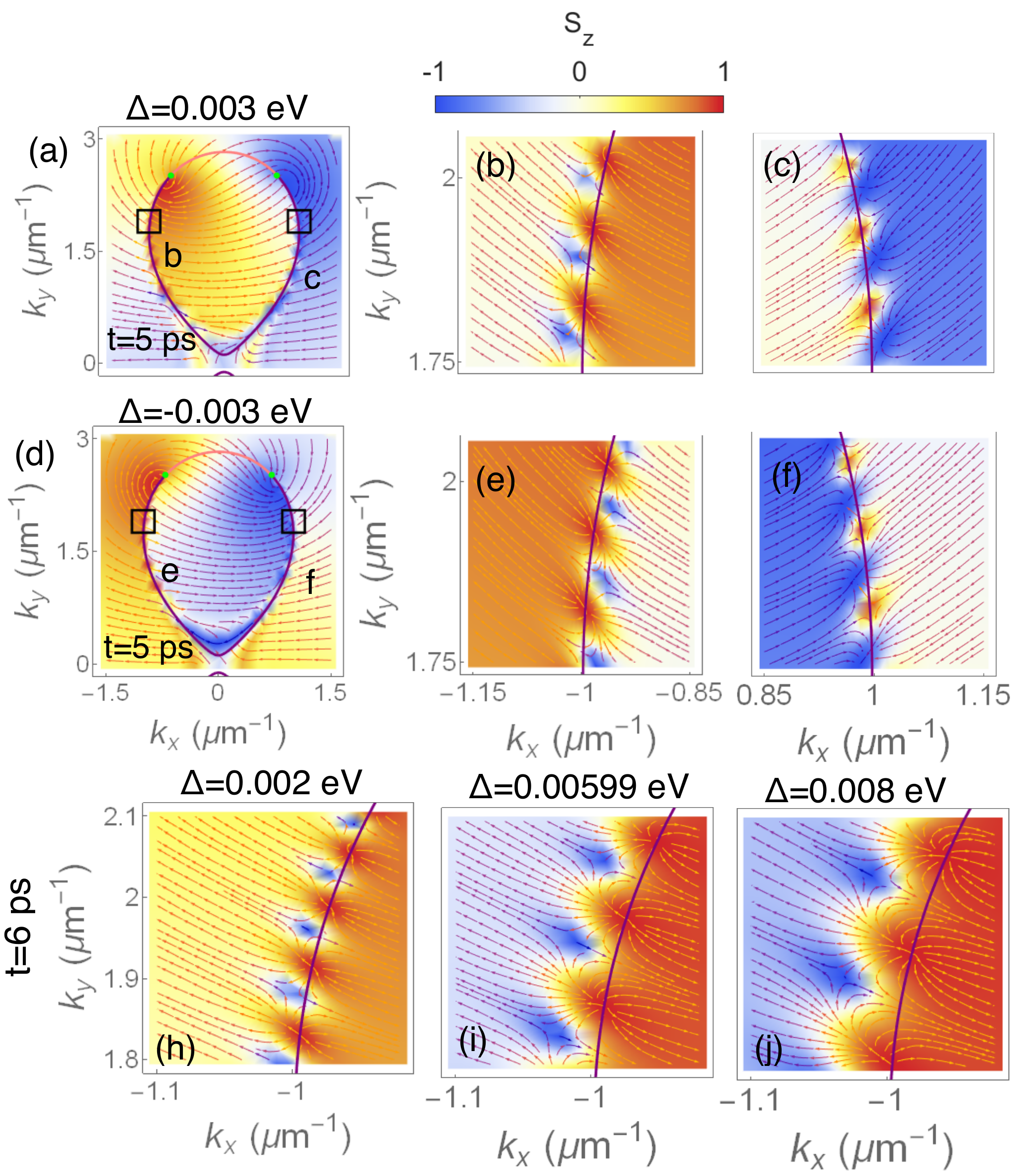}
    \caption{Pseudospin textures for (a-c) positive and (d-f) negative values of Zeeman splittings. Also showing the plots of the pseudospin defects in (h) a gapless phase with non-zero $\Delta$, (i) the phase where the two exceptional points merge and (j) a fully gapped phase. The wave packets used in these panels were initially circularly-polarized.}
    \label{fig: 8}
\end{figure}

Although the Zeeman splitting will not drastically change the in-plane textures $[S_x,S_y]$ of the defects, it has a huge impact on the $S_z$-textures of the eigenstates and the defects. As the strength of the Zeeman splitting increases, the pseudospin textures of one eigenstate will gain more positive $S_z$ contribution, while those of the other eigenstate will gain more negative $S_z$ contribution. As seen in Fig. \ref{fig: 8}, as $|\Delta|$ increases, $S_z$ becomes more negative on one side of the imaginary Fermi arc, and more positive on the other side. Consequently, as $|\Delta|$ increases, some of the defects will be slightly pushed away from the imaginary Fermi arcs. Which defects are pushed away from the imaginary Fermi arcs depends on the initial polarization of the exciton-polariton wave packet as well as the sign of the Zeeman splitting. For example, when the initial wave packet is in a circularly polarized mode, $\psi^+$, the core-down anti-merons and vortices are pushed away from the imaginary Fermi arcs, while the core-up defects stay on the arcs. As shown in Fig. \ref{fig: 8}(a-c), a positive Zeeman splitting pushes the core-down defects out of the regions bounded by Fermi arcs as they gain more negative $S_z$ in this region. Similarly, as seen in Fig. \ref{fig: 8}(d-f), a negative Zeeman splitting pushes the core-down defects into the regions bounded by Fermi arcs as the defects gain more negative $S_z$. If the wave packet is initially in the other circularly polarized mode, $\psi^-$, the opposite would happen, and the core-down defects would stay on the arc while the core-up defects would be pushed away.

Another consequence of the Zeeman splitting is that the core-up and core-down defects also gain more positive and negative $S_z$ contribution, respectively. If $|\Delta|$ continues to increase, eventually the pseudospin on the edge of the anti-merons will no longer lie on the $xy$-plane. Consequently, $S_z$ in these defects will no longer perform $\pi/2$ rotations with respects to the $xy$-plane, and they lose their anti-meron character as well. This is different from the self-acceleration, splitting and directional transport of the wave packets, which can persist at large $|\Delta|$.

\subsubsection{Pseudospin Defects in Real Space}\label{sec: defect r}
\begin{figure}[h]
    \centering
    \includegraphics[width=0.48\textwidth]{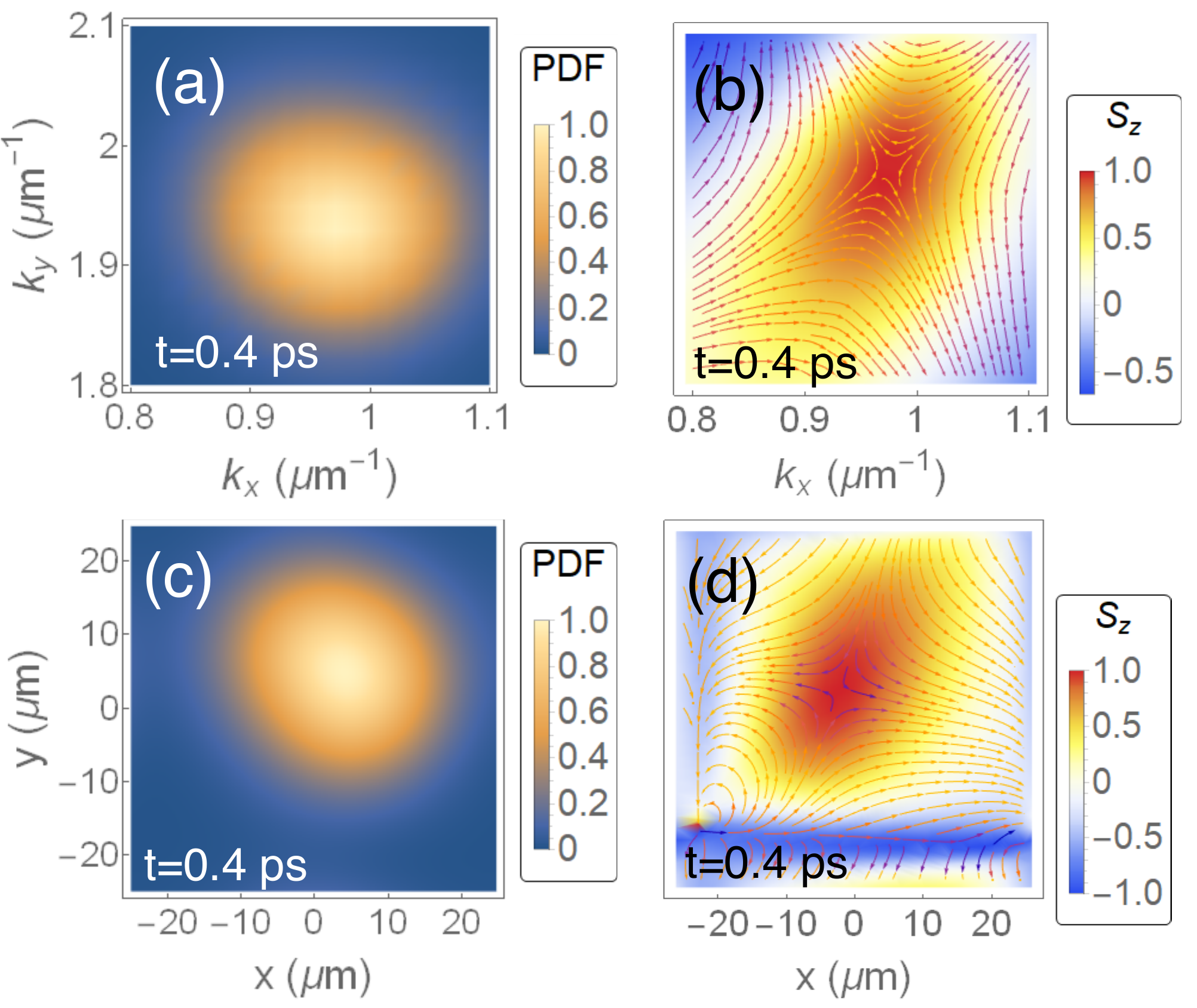}
    \caption{An exciton-polariton wave packet and its pseudospin textures in the momentum (a,b) and the real space (c,d).}
    \label{fig: 9}
\end{figure}

The pseudospin anti-merons on the imaginary Fermi arc in momentum space can also be measured in the real space. When the wave packet covers the anti-merons in momentum space, the anti-meron textures also emerge in the real space [see Fig. \ref{fig: 9}]. We should emphasise that the emergence of these anti-merons is a direct consequence of the frustration of pseudospins on the imaginary Fermi arc as the wave packet self-organizes into different eigenstates. This is in contrast to the previous studies of purely Hermitian two-dimensional systems, where the topological defects emerge due to the cavity anisotropy and TE-TM splitting \cite{cilibrizzi2016,vishnevsky2013,flayac2013,krol2021}. Our results show a new way to generate topological defects in the psuedospin structures that arise from the growth and decay of wave packets in a non-Hermitian system. These defects also represent a clear signature of the effects of the imaginary Fermi arc on the exciton-polariton dynamics in the real space.

Moreover, our preliminary estimates indicate that these defects can be observed experimentally in an exciton-polariton system based, e.g., on a perovskite embedded in a high quality microcavity \cite{su2021}, which would serve as a clear signature of the non-Hermitian effects on the wave packet dynamics (see Appendix \ref{sec: measure skyrmion}).

\section{Conclusion and outlook}
To conclude, we have investigated the dynamics of wave packets in a non-Hermitian exciton-polariton model. We found that the wave packets tend to self-organize into the eigenstate with the larger imaginary part of energy $\operatorname{Im}E$, then propagate towards the momentum with the corresponding $\max(\operatorname{Im}E)$. This results in splitting of the wave packets and a directional transport. We also observed core-up and core-down anti-merons on the imaginary Fermi arc in momentum space, which are destroyed by a sufficiently large Zeeman splitting.

However, there are still many open questions. It is yet unclear if the observables in the propagating wave packets, including the width, the centre-of-mass momentum and centre-of-mass positions, have simple analytic dependence on time, i.e., if they can be effectively described by a semi-classical equation of motion \cite{culcer2005,bleu2018wp,leblanc2021}. Furthermore, despite the fact that the trajectories of the exciton-polariton wave packets are aligned with the gradient of the loss rate in the system $\nabla\operatorname{Im}E$, this quantity does not always correctly describe the wave-packet trajectories under self-acceleration (see Appendix \ref{sec: 2 band WP}). The origin of the pseudospin defects and their connections to the Fermi arcs also require further investigation. 

Our results shed new light on the dynamics of wave packets and pseudospins in a non-Hermitian exciton-polariton system in momentum space as well as transport in real space. Furthermore, we believe that the emergent momentum space anti-meron textures in exciton-polariton pseudospins will enable the direct observation of the influence of the imaginary Fermi arcs on the dynamics of the system. All of the effects described above do not require an external potential and can be measured in an exciton-polariton system with optical anisotropy arising from the material properties of the semiconductor, e.g., perovskite \cite{su2021} and the organics \cite{liao2021}, or in cavity photon systems with an anisotropic spacer \cite{richter2019,krol2022}. Our results highlight the excellent potential of exciton polaritons as a platform to study non-Hermitian dynamics.

\begin{acknowledgments}
We acknowledge support from the Australian Research Council (ARC) through the Centre of Excellence Grant CE170100039 and the Discovery Early Career Researcher Award DE220100712, and Australian Government Research Training Program (RTP) Scholarship.
\end{acknowledgments}

\appendix

\section{Values of Parameters}\label{sec: para}
For the figures presented in this work, we chose the parameters such that the exceptional points are well separated and the bulk Fermi arcs can be clearly seen by the readers. We set $\chi=3.75\times10^{-4}$ $\mu$m$^4$eV, and choose the mean energy, linewidth and effective polariton mass to be $E_0-i\gamma_0=2.306-4.5\times10^{-4}i$ eV and $\hbar^2/2m\approx2.3\times10^{-3}$ $\mu$m$^2$eV. We then set the optical anisotropy and the photonic spin-orbit coupling to be $\alpha=8\times10^{-3}$ eV and $\beta=10^{-3}$ $\mu$m$^{2}$eV. We then set the photonic losses to be $a=10^{-5}$ eV, $b=7.5\times10^{-4}$ $\mu$m$^{2}$eV and assume there is no Zeeman splitting, $\Delta=0$ eV. Using this set of parameters, the exceptional points have momenta of $k\approx 2.5$ $\mu$m$^{-1}$ and each pair is separated by $\approx 1.6$ $\mu$m$^{-1}$ in momentum space. In a realistic system such as the perovskite-based exciton polaritons described in Ref. \cite{su2021}, the exceptional points would have momenta of $k\approx 6.7$ $\mu$m$^{-1}$ and each pair is separated by $\approx 0.24$ $\mu$m$^{-1}$ in momentum space, which would be difficult for the readers to observe.

\section{Time-Evolution Operator in Exciton-Polariton Model}\label{sec: evolution}
We consider a general two-band non-Hermitian Hamiltonian of the form
$$\mathbf{H}(\mathbf{k})={H}_0(\mathbf{k}){\mathbf{I}} + \overrightarrow{\mathbf{G}}(\mathbf{k})\cdot\overrightarrow{\sigma},$$
where $\overrightarrow{\mathbf{G}}(\mathbf{k})=[G_x(\mathbf{k}),G_y(\mathbf{k}),G_z(\mathbf{k})]$ is the gauge field, $\overrightarrow{\sigma}$ is a vector of Pauli matrices, and ${\mathbf{I}}$ is the $2\times 2$ identity matrix. Dropping the $\mathbf{k}$ dependence for brevity, the time-evolution operator can then be written as
\begin{equation}
e^{-i\mathbf{H}t/\hbar}=e^{-i{H_0}t/\hbar} \left( {\mathbf{I}} \cos{\frac{Gt}{\hbar}} - i \frac{\overrightarrow{\mathbf{G}}}{G}\cdot\overrightarrow{\sigma} \sin{\frac{Gt}{\hbar}} \right),
\end{equation}
where $G = \sqrt{G_x^2+G_y^2+G_z^2}$. Note that the complex energy spectrum can be written simply as $E_\pm = H_0 \pm G$.

We should also mention that this formalism does not hold at the exceptional points where the Hamiltonian can no longer be diagonalized and the two eigenstates will coalesce. However, since this formalism holds in the vicinity of the exceptional points as well as for the rest of momentum space, we believe that it is an accurate approximation for the time-evolution of the exciton-polariton wave packets.

\section{Overlap with Eigenstates}\label{sec: eigenstates}
In a quantum system described by a non-Hermitian Hamiltonian operator, the right and left eigenstates are generally not equal and can be written as \cite{ghatak2019,bergholz2021}:
\begin{equation}
    \begin{split}
        \mathbf{H}|\psi^R_\pm\rangle&=E_\pm|\psi_\pm^R\rangle\\
        \mathbf{H}^\dagger|\psi^L_\pm\rangle&=E^*_\pm|\psi_\pm^L\rangle.
    \end{split}
\end{equation}
Instead of the usual orthonormal condition, the eigenstates obey the biorthogonal condition
\begin{equation}
    \langle\psi_i^L|\psi_j^R\rangle=\delta_{i,j}.
\end{equation}
Therefore, when the system is in a superposition of the two right eigenstates $|\psi\rangle=c_+|\psi_+^R\rangle+|\psi_-^R\rangle$, the coefficients can be determined as $$c_\pm=\langle\psi_\pm^L|\psi\rangle.$$
By taking the normalized modulus squared $|c_\pm|^2/(|c_+|^2+|c_-|^2)$, we were able to calculate the ratio of each eigenstate in the wave packet. Similarly to the time-evolution operator in Appendix \ref{sec: evolution}, this approach fail at the exceptional points where the two eigenstates coalesce. Therefore, we believe that the approximation will be accurate as it will hold in the vicinity of the exceptional points.

\section{Exciton-Polariton Pseudospins}\label{sec: S}
The pseudospins of the exciton polaritons can be calculated from the polarization components as
\begin{equation}\label{eq: pseudospin}
    \begin{split}
        S_x&=\frac{|\psi^H|^2-|\psi^V|^2}{|\psi^H|^2+|\psi^V|^2}\\
        S_y&=\frac{|\psi^D|^2-|\psi^A|^2}{|\psi^D|^2+|\psi^A|^2}\\
        S_z&=\frac{|\psi^+|^2-|\psi^-|^2}{|\psi^+|^2+|\psi^-|^2}.
    \end{split}
\end{equation}
Here, $|\psi^{H,V}|^2$ represents the horizontal and vertical polarization intensities, $|\psi^{D,A}|^2$ represents the diagonal and anti-diagonal polarization intensities and $|\psi^{\pm}|^2$ represents the circular polarization intensities \cite{bleu2018}.

\section{Proposed Experimental Measurement of the Pseudospin Defects}\label{sec: measure skyrmion}
In experiments, the probability density distribution and the pseudospin textures can be measured using a streak camera which can take a screen shot of the wave packets in both real and momentum space at a resolution of $\approx2$ ps. Since the measurements are made by taking multiple screen shots and then averaging over time resolution, we aim to investigate whether the pseudospin defects on the imaginary Fermi arc will still appear when the fields are averaged. 

\begin{figure}[h]
    \centering
    \includegraphics[width=0.48\textwidth]{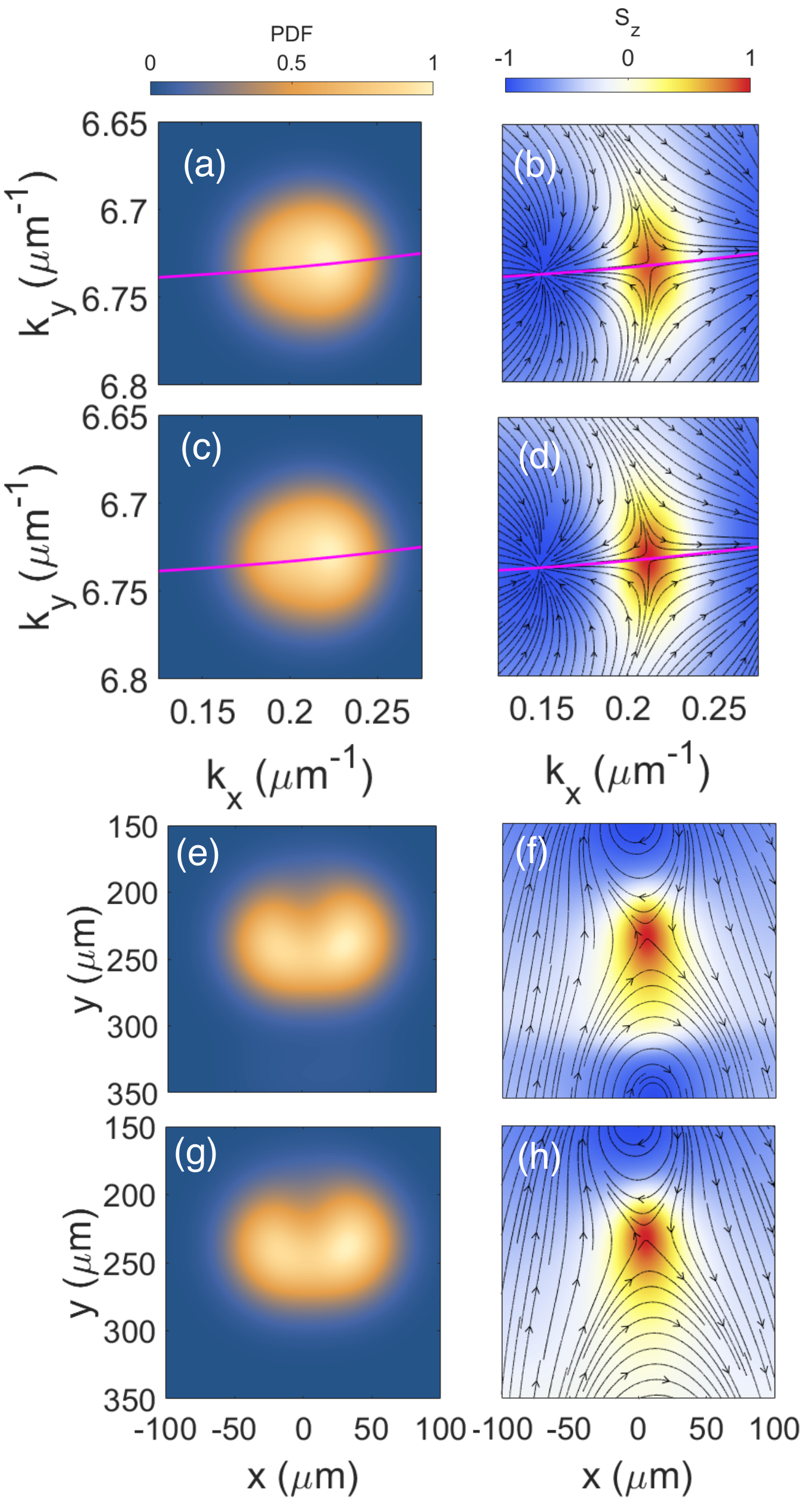}
    \caption{The averaged wave packet and pseudospin textures of the exciton-polariton in momentum space (a,b) and in the real space (e,f). The probability density distribution and the pseudospin take similar forms to the wave packet and pseudospin textures at $t=5$ ps in (c,d) and (g,h). The purple line denotes the imaginary Fermi arc.}
    \label{fig: appendix 1}
\end{figure}

Since the pseudospin of the wave packets cannot be measured in the regions with small probability density, we need to make sure that the exciton-polariton density at the location of the pseudospin defects is non-negligible. Previous work \cite{solnyshkov2021} shows that the larger the wave packets are, the faster they will move away from the initial centre-of-mass momentum. Therefore, we need a wave packet that is well localized in momentum space, so it will stay near the defect for a long time. However, the wave packet cannot be too focused since we still need to measure the pseudospin texture of the entire defect. 

Furthermore, although it is unknown whether the change in the pseudospin textures will reach a steady state, our numerical calculations indicate that the change of the pseudospin slows down with time. Therefore, the best way to measure the pseudospin defects is to take measurements over a long time, taking care to not exceed the time after which the wave packet moves away from the imaginary Fermi arc.

To simulate an experiment, we choose a circularly-polarized exciton-polariton wave packet with a width of $\sigma_k=0.02$ $\mu$m$^{-1}$ in momentum space with initial centre-of-mass momentum of $\langle\mathbf{k}\rangle=(0.215,6.73)$ $\mu$m$^{-1}$ and considered a realistic set of parameters presented in Ref. \cite{su2021}. The initial condition of the wave packet is chosen so that its size is small enough so it will stay on the imaginary Fermi arc for a long time, yet not too small to be realized in experiment. The wave packet has a size of $\sigma_r=50$ $\mu$m in real space, which can be achieved in the experiments. We found that that the wave packet at $t=5$ ps is still on the imaginary Fermi arc and covers the pseudospin defect, but is starting to move away at $t=15$ ps. Therefore, we chose to take the average of the wave packets every $2$ ps from $t=5$ ps to $t=15$ ps to see if the averaged wave packet will exhibit the pseudospin defects.

The polarizations, $|\psi^\pm|^2$, $|\psi^{H,V}|^2$ and $|\psi^{D,A}|^2$ are therefore measured every $2$ ps from $t=5$ ps to $t=15$ ps. We then take the average of these intensities $\langle|\psi^\pm|^2\rangle_{ave}$, $\langle|\psi^{H,V}|^2\rangle_{ave}$, $\langle|\psi^{D,A}|^2\rangle_{ave}$ and use them to compute the pseudopin textures. We observe the wave packet and its time-averaged pseudospin textures take forms similar to those at $t=5$ ps in momentum space [see Fig. \ref{fig: appendix 1}(a-d)]. Therefore the snapshot of the wave packet at $t=5$ ps will dominate the time-averaged measurement in momentum space. This is because the wave packets in later times have much lower intensities due to the continuing decay (negative imaginary part of eigenenergies) of the exciton polaritons. On the other hand, in the real space, the averaged wave packet and pseudospin textures do not completely agree with those at $t=5$ ps as seen in Fig. \ref{fig: appendix 1}(e-h). However, the pseudospin anti-meron structure at the centre of the wave packet still persist after the averaging. Therefore, we conclude that the pseudospin anti-merons generated by non-Hermitian wave packet dynamics can be measured experimentally in both momentum space and real space.

\section{Wave-Packet Trajectories in Other Two-Band Systems}\label{sec: 2 band WP}
Previous studies \cite{solnyshkov2021} have suggested that the wave-packet centre-of-mass momenta of a non-Hermitian Dirac model follow the gradient of the imaginary part of the eigenenergies $\nabla\operatorname{Im}E$. Although this seems to accurately describe the trajectories of the exciton-polariton wave packets, this is not always the case. In this section, we illustrate this point by presenting the analysis of the wave packet dynamics in two exemplary models, a non-Hermitian Dirac model and a non-Hermitian Chern insulator. These two models have similar band structures near the origin, but their wave-packet dynamics and asymptotic behaviours are qualitatively different.

\begin{figure}[h]
    \centering
    \includegraphics[width=0.48\textwidth]{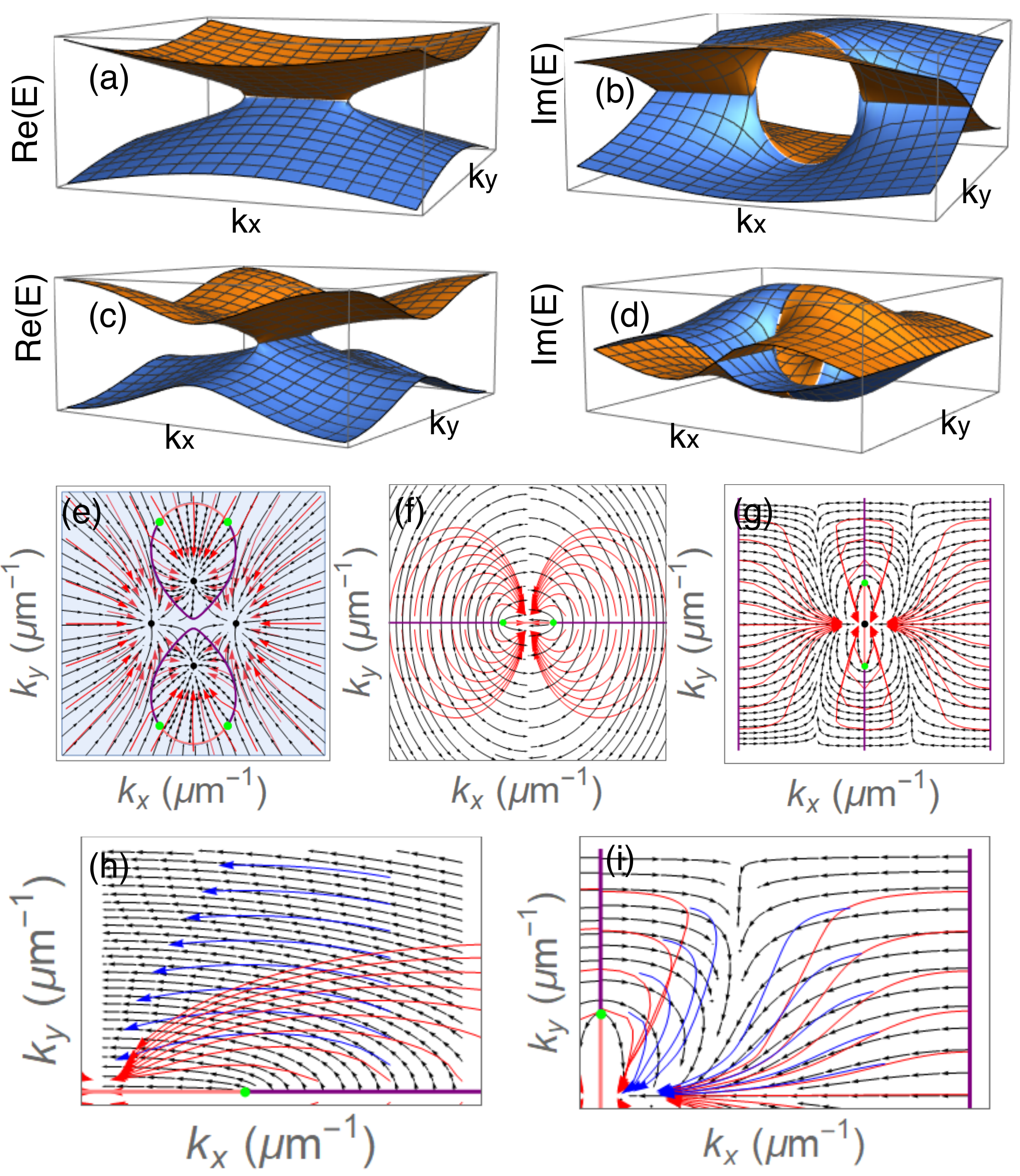}
    \caption{(a), (b): The real and imaginary parts of the eigenenergies of the non-Hermitian Dirac model.
    (c), (d): The real and imaginary parts of the eigenenergies of the non-Hermitian Chern insulator in the nodal phase. Also showing the wave-packet trajectories (red) with $\nabla\operatorname{Im}E$ (black) with the Fermi arcs and exceptional points of the (e) exciton-polariton model, (f) the non-Hermitian Dirac model and (g) the non-Hermitian Chern insulator. (h), (i): Wave-packet trajectories with different sets of initial centre-of-mass momenta (blue arrows) in non-Hermitian Dirac model and non-Hermitian Chern insulator, respectively, showing that they do not follow the same paths indicated the red arrows.}
    \label{fig: appendix 2}
\end{figure}

\subsection{Non-Hermitian Dirac Model}
We considered the Dirac model, described by the Hamiltonian
\begin{equation}
    \begin{split}
        \mathbf{H}&=\hbar ck_x\sigma_x+\hbar ck_y\sigma_y.
    \end{split}
\end{equation}
This model has eigenenergies of $E_\pm=\pm\hbar c\sqrt{k_x^2+k_y^2}$ and has a spectral degeneracy at the origin called the Dirac point. In solid state physics, this model describes the Dirac semimetal, which has elementary excitations of massless Dirac fermions and describes the electronic properties of graphenes \cite{novoselov2004,novoselov2005,zhang2005,castro2009,armitage2018}. It also plays an important role in optical physics. When it is written in the basis of circularly-polarized modes, the Dirac point describes a diabolical point which exhibits conical refraction \cite{tercas2014,solnyshkov2021,hamilton1837,lloyd1837}.

The non-Hermitian Dirac model that we considered is described by the following Hamiltonian \cite{solnyshkov2021}
\begin{equation}
    \mathbf{H}=\hbar c k_x\sigma_x+(\hbar ck_y+i\kappa)\sigma_y+\Delta\sigma_z,
\end{equation}
which has the eigenenergies
\begin{equation}\label{eq: NH Dirac E}
    E_\pm=\pm\sqrt{\hbar^2c^2k_x^2+(\hbar ck_y+i\kappa)^2+\Delta^2}.
\end{equation}
The non-Hermitian term $i\kappa\sigma_y$ splits the Dirac point at $\mathbf{k}^{DP}=(0,0)$ into a pair of non-Hermitian degeneracies called the exceptional points in momentum space at $\mathbf{k}^{EP}=(\pm\sqrt{\kappa^2-\Delta^2}/(\hbar c),0)$.

In non-Hermitian Dirac model, $E_+$ has the larger imaginary part at $k_y<0$ while $E_-$ has the larger imaginary part at $k_y>0$ [see Fig. \ref{fig: appendix 2}(a,b)]. Therefore, as time evolves, the wave packets at $k_y<0$ tend to self-organize into the upper eigenstate, while the wave packets at $k_y>0$ tend to self-organize into the lower eigenstate. Furthermore, at $\Delta=0$, $\max(\operatorname{Im}E)$ lies at the $k_y$-axis. Therefore, the wave packets in the non-Hermitian Dirac model propagate towards to a line of fixed points, unlike the wave packets in the exciton-polariton model. From Fig. \ref{fig: appendix 2}(e,f), we can observe that, unlike the trajectories of the exciton-polariton wave packets, the centre-of-mass momenta of the Dirac wave packets seem to only follow $\nabla\operatorname{Im}E_\pm$ initially. As time evolves, the trajectories of the wave packets start to deviate from $\nabla\operatorname{Im}E_\pm$. More importantly, we can see that the trajectories of the wave packets as well as the fixed points they propagate to highly depend on their initial momenta. From Fig. \ref{fig: appendix 2}(h), we can see that the trajectories of the wave packets with different set of initial centre-of-mass momenta will not follow the same paths towards the same fixed points as the wave packets in Fig. \ref{fig: appendix 2}(f).
\\
\subsection{Non-Hermitian Chern Insulator}
The non-Hermitian Chern insulator is described by the Hamiltonian \cite{kawabata2018}
\begin{equation}
    \mathbf{H}=(m+\cos k_x+\cos k_y)\sigma_x+(i\gamma+\sin k_x)\sigma_y+\sin k_y\sigma_z
\end{equation}
and has eigenenergies
\begin{equation}
    E_\pm=\pm\Big((m+\cos k_x+\cos k_y)^2+(i\gamma+\sin k_x)^2+\sin^2 k_y\Big)^{1/2}.
\end{equation} 
The phase diagram and edge states of this model have been extensively studied in Ref. \cite{kawabata2018}. At $m=-2$, $\gamma=1$, there is a pair of exceptional points in the Brillouine zone at $\mathbf{k}=(0,\pm\pi/3)$ and the band structure is similar to that of the non-Hermitian Dirac model. However, unlike the Dirac model, there is a maximum of $\operatorname{Im}E_\pm$ at the middle of the Brillouine zone, which is also the middle of the bulk Fermi arc [see Fig. \ref{fig: appendix 2}(c,d)].

The wave packets with different initial centre-of-mass momenta in this model also follow different paths [see Fig. \ref{fig: appendix 2}(g)]. However, unlike the non-Hermitian Dirac model, where the asymptotic behaviours of the wave-packet centre-of-mass momenta are sensitive to the initial conditions, the wave packets in the non-Hermitian Chern insulator seem to propagate to the same fixed point asymptotically.

The above analysis of the non-Hermitian Dirac model and the non-Hermitian Chern insulator shows that the centre-of-mass momenta of a wave packet only follow $\nabla\operatorname{Im}E$ of its corresponding eigenstate on a short time scale. The trajectories of the wave packets will deviate increasingly from $\nabla\operatorname{Im}E$ as time evolves. Furthermore, in cases where there are multiple maxima of $\operatorname{Im}E_\pm$, both the paths that the wave packets take and the fixed points that the wave packets propagate towards to will be sensitive to the initial conditions of the wave packets.


\bibliography{refs}

\begin{thebibliography}{50}%
\makeatletter
\providecommand \@ifxundefined [1]{%
 \@ifx{#1\undefined}
}%
\providecommand \@ifnum [1]{%
 \ifnum #1\expandafter \@firstoftwo
 \else \expandafter \@secondoftwo
 \fi
}%
\providecommand \@ifx [1]{%
 \ifx #1\expandafter \@firstoftwo
 \else \expandafter \@secondoftwo
 \fi
}%
\providecommand \natexlab [1]{#1}%
\providecommand \enquote  [1]{``#1''}%
\providecommand \bibnamefont  [1]{#1}%
\providecommand \bibfnamefont [1]{#1}%
\providecommand \citenamefont [1]{#1}%
\providecommand \href@noop [0]{\@secondoftwo}%
\providecommand \href [0]{\begingroup \@sanitize@url \@href}%
\providecommand \@href[1]{\@@startlink{#1}\@@href}%
\providecommand \@@href[1]{\endgroup#1\@@endlink}%
\providecommand \@sanitize@url [0]{\catcode `\\12\catcode `\$12\catcode
  `\&12\catcode `\#12\catcode `\^12\catcode `\_12\catcode `\%12\relax}%
\providecommand \@@startlink[1]{}%
\providecommand \@@endlink[0]{}%
\providecommand \url  [0]{\begingroup\@sanitize@url \@url }%
\providecommand \@url [1]{\endgroup\@href {#1}{\urlprefix }}%
\providecommand \urlprefix  [0]{URL }%
\providecommand \Eprint [0]{\href }%
\providecommand \doibase [0]{https://doi.org/}%
\providecommand \selectlanguage [0]{\@gobble}%
\providecommand \bibinfo  [0]{\@secondoftwo}%
\providecommand \bibfield  [0]{\@secondoftwo}%
\providecommand \translation [1]{[#1]}%
\providecommand \BibitemOpen [0]{}%
\providecommand \bibitemStop [0]{}%
\providecommand \bibitemNoStop [0]{.\EOS\space}%
\providecommand \EOS [0]{\spacefactor3000\relax}%
\providecommand \BibitemShut  [1]{\csname bibitem#1\endcsname}%
\let\auto@bib@innerbib\@empty
\bibitem [{\citenamefont {Ghatak}\ and\ \citenamefont
  {Das}(2019)}]{ghatak2019}%
  \BibitemOpen
  \bibfield  {author} {\bibinfo {author} {\bibfnamefont {A.}~\bibnamefont
  {Ghatak}}\ and\ \bibinfo {author} {\bibfnamefont {T.}~\bibnamefont {Das}},\
  }\bibfield  {title} {\bibinfo {title} {New topological invariants in
  non-{H}ermitian systems},\ }\href {https://doi.org/10.1088/1361-648x/ab11b3}
  {\bibfield  {journal} {\bibinfo  {journal} {Journal of Physics: Condensed
  Matter}\ }\textbf {\bibinfo {volume} {31}},\ \bibinfo {pages} {263001}
  (\bibinfo {year} {2019})}\BibitemShut {NoStop}%
\bibitem [{\citenamefont {Bergholtz}\ \emph {et~al.}(2021)\citenamefont
  {Bergholtz}, \citenamefont {Budich},\ and\ \citenamefont
  {Kunst}}]{bergholz2021}%
  \BibitemOpen
  \bibfield  {author} {\bibinfo {author} {\bibfnamefont {E.~J.}\ \bibnamefont
  {Bergholtz}}, \bibinfo {author} {\bibfnamefont {J.~C.}\ \bibnamefont
  {Budich}},\ and\ \bibinfo {author} {\bibfnamefont {F.~K.}\ \bibnamefont
  {Kunst}},\ }\bibfield  {title} {\bibinfo {title} {Exceptional topology of
  non-{H}ermitian systems},\ }\href
  {https://doi.org/10.1103/RevModPhys.93.015005} {\bibfield  {journal}
  {\bibinfo  {journal} {Rev. Mod. Phys.}\ }\textbf {\bibinfo {volume} {93}},\
  \bibinfo {pages} {015005} (\bibinfo {year} {2021})}\BibitemShut {NoStop}%
\bibitem [{\citenamefont {El-Ganainy}\ \emph {et~al.}(2018)\citenamefont
  {El-Ganainy}, \citenamefont {Makris}, \citenamefont {Khajavikhan},
  \citenamefont {Musslimani}, \citenamefont {Rotter},\ and\ \citenamefont
  {Christodoulides}}]{el-ganainy2018}%
  \BibitemOpen
  \bibfield  {author} {\bibinfo {author} {\bibfnamefont {R.}~\bibnamefont
  {El-Ganainy}}, \bibinfo {author} {\bibfnamefont {K.~G.}\ \bibnamefont
  {Makris}}, \bibinfo {author} {\bibfnamefont {M.}~\bibnamefont {Khajavikhan}},
  \bibinfo {author} {\bibfnamefont {Z.~H.}\ \bibnamefont {Musslimani}},
  \bibinfo {author} {\bibfnamefont {S.}~\bibnamefont {Rotter}},\ and\ \bibinfo
  {author} {\bibfnamefont {D.~N.}\ \bibnamefont {Christodoulides}},\ }\bibfield
   {title} {\bibinfo {title} {Non-{H}ermitian physics and {PT} symmetry},\
  }\href {https://doi.org/10.1038/nphys4323} {\bibfield  {journal} {\bibinfo
  {journal} {Nature Physics}\ }\textbf {\bibinfo {volume} {14}},\ \bibinfo
  {pages} {11} (\bibinfo {year} {2018})}\BibitemShut {NoStop}%
\bibitem [{\citenamefont {{\"O}zdemir}\ \emph {et~al.}(2019)\citenamefont
  {{\"O}zdemir}, \citenamefont {Rotter}, \citenamefont {Nori},\ and\
  \citenamefont {Yang}}]{ozdemir2019}%
  \BibitemOpen
  \bibfield  {author} {\bibinfo {author} {\bibfnamefont {c.~K.}\ \bibnamefont
  {{\"O}zdemir}}, \bibinfo {author} {\bibfnamefont {S.}~\bibnamefont {Rotter}},
  \bibinfo {author} {\bibfnamefont {F.}~\bibnamefont {Nori}},\ and\ \bibinfo
  {author} {\bibfnamefont {L.}~\bibnamefont {Yang}},\ }\bibfield  {title}
  {\bibinfo {title} {Parity--time symmetry and exceptional points in
  photonics},\ }\href {https://doi.org/10.1038/s41563-019-0304-9} {\bibfield
  {journal} {\bibinfo  {journal} {Nature Materials}\ }\textbf {\bibinfo
  {volume} {18}},\ \bibinfo {pages} {783} (\bibinfo {year} {2019})}\BibitemShut
  {NoStop}%
\bibitem [{\citenamefont {Gao}\ \emph {et~al.}(2015)\citenamefont {Gao},
  \citenamefont {Estrecho}, \citenamefont {Bliokh}, \citenamefont {Liew},
  \citenamefont {Fraser}, \citenamefont {Brodbeck}, \citenamefont {Kamp},
  \citenamefont {Schneider}, \citenamefont {H{\"o}fling}, \citenamefont
  {Yamamoto}, \citenamefont {Nori}, \citenamefont {Kivshar}, \citenamefont
  {Truscott}, \citenamefont {Dall},\ and\ \citenamefont
  {Ostrovskaya}}]{gao2015}%
  \BibitemOpen
  \bibfield  {author} {\bibinfo {author} {\bibfnamefont {T.}~\bibnamefont
  {Gao}}, \bibinfo {author} {\bibfnamefont {E.}~\bibnamefont {Estrecho}},
  \bibinfo {author} {\bibfnamefont {K.~Y.}\ \bibnamefont {Bliokh}}, \bibinfo
  {author} {\bibfnamefont {T.~C.~H.}\ \bibnamefont {Liew}}, \bibinfo {author}
  {\bibfnamefont {M.~D.}\ \bibnamefont {Fraser}}, \bibinfo {author}
  {\bibfnamefont {S.}~\bibnamefont {Brodbeck}}, \bibinfo {author}
  {\bibfnamefont {M.}~\bibnamefont {Kamp}}, \bibinfo {author} {\bibfnamefont
  {C.}~\bibnamefont {Schneider}}, \bibinfo {author} {\bibfnamefont
  {S.}~\bibnamefont {H{\"o}fling}}, \bibinfo {author} {\bibfnamefont
  {Y.}~\bibnamefont {Yamamoto}}, \bibinfo {author} {\bibfnamefont
  {F.}~\bibnamefont {Nori}}, \bibinfo {author} {\bibfnamefont {Y.~S.}\
  \bibnamefont {Kivshar}}, \bibinfo {author} {\bibfnamefont {A.~G.}\
  \bibnamefont {Truscott}}, \bibinfo {author} {\bibfnamefont {R.~G.}\
  \bibnamefont {Dall}},\ and\ \bibinfo {author} {\bibfnamefont {E.~A.}\
  \bibnamefont {Ostrovskaya}},\ }\bibfield  {title} {\bibinfo {title}
  {Observation of non-{H}ermitian degeneracies in a chaotic exciton-polariton
  billiard},\ }\href {https://doi.org/10.1038/nature15522} {\bibfield
  {journal} {\bibinfo  {journal} {Nature}\ }\textbf {\bibinfo {volume} {526}},\
  \bibinfo {pages} {554} (\bibinfo {year} {2015})}\BibitemShut {NoStop}%
\bibitem [{\citenamefont {Su}\ \emph {et~al.}(2021)\citenamefont {Su},
  \citenamefont {Estrecho}, \citenamefont {Biega{\'n}ska}, \citenamefont
  {Huang}, \citenamefont {Wurdack}, \citenamefont {Pieczarka}, \citenamefont
  {Truscott}, \citenamefont {Liew}, \citenamefont {Ostrovskaya},\ and\
  \citenamefont {Xiong}}]{su2021}%
  \BibitemOpen
  \bibfield  {author} {\bibinfo {author} {\bibfnamefont {R.}~\bibnamefont
  {Su}}, \bibinfo {author} {\bibfnamefont {E.}~\bibnamefont {Estrecho}},
  \bibinfo {author} {\bibfnamefont {D.}~\bibnamefont {Biega{\'n}ska}}, \bibinfo
  {author} {\bibfnamefont {Y.}~\bibnamefont {Huang}}, \bibinfo {author}
  {\bibfnamefont {M.}~\bibnamefont {Wurdack}}, \bibinfo {author} {\bibfnamefont
  {M.}~\bibnamefont {Pieczarka}}, \bibinfo {author} {\bibfnamefont {A.~G.}\
  \bibnamefont {Truscott}}, \bibinfo {author} {\bibfnamefont {T.~C.~H.}\
  \bibnamefont {Liew}}, \bibinfo {author} {\bibfnamefont {E.~A.}\ \bibnamefont
  {Ostrovskaya}},\ and\ \bibinfo {author} {\bibfnamefont {Q.}~\bibnamefont
  {Xiong}},\ }\bibfield  {title} {\bibinfo {title} {Direct measurement of a
  non-{H}ermitian topological invariant in a hybrid light-matter system},\
  }\href {https://doi.org/10.1126/sciadv.abj8905} {\bibfield  {journal}
  {\bibinfo  {journal} {Science Advances}\ }\textbf {\bibinfo {volume} {7}},\
  \bibinfo {pages} {eabj8905} (\bibinfo {year} {2021})},\ \Eprint
  {https://arxiv.org/abs/https://www.science.org/doi/pdf/10.1126/sciadv.abj8905}
  {https://www.science.org/doi/pdf/10.1126/sciadv.abj8905} \BibitemShut
  {NoStop}%
\bibitem [{\citenamefont {Kunst}\ \emph {et~al.}(2018)\citenamefont {Kunst},
  \citenamefont {Edvardsson}, \citenamefont {Budich},\ and\ \citenamefont
  {Bergholtz}}]{kunst2018}%
  \BibitemOpen
  \bibfield  {author} {\bibinfo {author} {\bibfnamefont {F.~K.}\ \bibnamefont
  {Kunst}}, \bibinfo {author} {\bibfnamefont {E.}~\bibnamefont {Edvardsson}},
  \bibinfo {author} {\bibfnamefont {J.~C.}\ \bibnamefont {Budich}},\ and\
  \bibinfo {author} {\bibfnamefont {E.~J.}\ \bibnamefont {Bergholtz}},\
  }\bibfield  {title} {\bibinfo {title} {Biorthogonal bulk-boundary
  correspondence in non-{H}ermitian systems},\ }\href
  {https://doi.org/10.1103/PhysRevLett.121.026808} {\bibfield  {journal}
  {\bibinfo  {journal} {Phys. Rev. Lett.}\ }\textbf {\bibinfo {volume} {121}},\
  \bibinfo {pages} {026808} (\bibinfo {year} {2018})}\BibitemShut {NoStop}%
\bibitem [{\citenamefont {Hofmann}\ \emph {et~al.}(2020)\citenamefont
  {Hofmann}, \citenamefont {Helbig}, \citenamefont {Schindler}, \citenamefont
  {Salgo}, \citenamefont {Brzezi\ifmmode~\acute{n}\else \'{n}\fi{}ska},
  \citenamefont {Greiter}, \citenamefont {Kiessling}, \citenamefont {Wolf},
  \citenamefont {Vollhardt}, \citenamefont {Kaba\ifmmode~\check{s}\else
  \v{s}\fi{}i}, \citenamefont {Lee}, \citenamefont {Bilu\ifmmode \check{s}\else
  \v{s}\fi{}i\ifmmode~\acute{c}\else \'{c}\fi{}}, \citenamefont {Thomale},\
  and\ \citenamefont {Neupert}}]{hofmann2020}%
  \BibitemOpen
  \bibfield  {author} {\bibinfo {author} {\bibfnamefont {T.}~\bibnamefont
  {Hofmann}}, \bibinfo {author} {\bibfnamefont {T.}~\bibnamefont {Helbig}},
  \bibinfo {author} {\bibfnamefont {F.}~\bibnamefont {Schindler}}, \bibinfo
  {author} {\bibfnamefont {N.}~\bibnamefont {Salgo}}, \bibinfo {author}
  {\bibfnamefont {M.}~\bibnamefont {Brzezi\ifmmode~\acute{n}\else
  \'{n}\fi{}ska}}, \bibinfo {author} {\bibfnamefont {M.}~\bibnamefont
  {Greiter}}, \bibinfo {author} {\bibfnamefont {T.}~\bibnamefont {Kiessling}},
  \bibinfo {author} {\bibfnamefont {D.}~\bibnamefont {Wolf}}, \bibinfo {author}
  {\bibfnamefont {A.}~\bibnamefont {Vollhardt}}, \bibinfo {author}
  {\bibfnamefont {A.}~\bibnamefont {Kaba\ifmmode~\check{s}\else \v{s}\fi{}i}},
  \bibinfo {author} {\bibfnamefont {C.~H.}\ \bibnamefont {Lee}}, \bibinfo
  {author} {\bibfnamefont {A.}~\bibnamefont {Bilu\ifmmode \check{s}\else
  \v{s}\fi{}i\ifmmode~\acute{c}\else \'{c}\fi{}}}, \bibinfo {author}
  {\bibfnamefont {R.}~\bibnamefont {Thomale}},\ and\ \bibinfo {author}
  {\bibfnamefont {T.}~\bibnamefont {Neupert}},\ }\bibfield  {title} {\bibinfo
  {title} {Reciprocal skin effect and its realization in a topolectrical
  circuit},\ }\href {https://doi.org/10.1103/PhysRevResearch.2.023265}
  {\bibfield  {journal} {\bibinfo  {journal} {Phys. Rev. Research}\ }\textbf
  {\bibinfo {volume} {2}},\ \bibinfo {pages} {023265} (\bibinfo {year}
  {2020})}\BibitemShut {NoStop}%
\bibitem [{\citenamefont {Weidemann}\ \emph {et~al.}(2020)\citenamefont
  {Weidemann}, \citenamefont {Kremer}, \citenamefont {Helbig}, \citenamefont
  {Hofmann}, \citenamefont {Stegmaier}, \citenamefont {Greiter}, \citenamefont
  {Thomale},\ and\ \citenamefont {Szameit}}]{weidemann2020}%
  \BibitemOpen
  \bibfield  {author} {\bibinfo {author} {\bibfnamefont {S.}~\bibnamefont
  {Weidemann}}, \bibinfo {author} {\bibfnamefont {M.}~\bibnamefont {Kremer}},
  \bibinfo {author} {\bibfnamefont {T.}~\bibnamefont {Helbig}}, \bibinfo
  {author} {\bibfnamefont {T.}~\bibnamefont {Hofmann}}, \bibinfo {author}
  {\bibfnamefont {A.}~\bibnamefont {Stegmaier}}, \bibinfo {author}
  {\bibfnamefont {M.}~\bibnamefont {Greiter}}, \bibinfo {author} {\bibfnamefont
  {R.}~\bibnamefont {Thomale}},\ and\ \bibinfo {author} {\bibfnamefont
  {A.}~\bibnamefont {Szameit}},\ }\bibfield  {title} {\bibinfo {title}
  {Topological funneling of light},\ }\href@noop {} {\bibfield  {journal}
  {\bibinfo  {journal} {Science}\ }\textbf {\bibinfo {volume} {368}},\ \bibinfo
  {pages} {311 } (\bibinfo {year} {2020})}\BibitemShut {NoStop}%
\bibitem [{\citenamefont {Jin}\ and\ \citenamefont {Song}(2018)}]{jin2018}%
  \BibitemOpen
  \bibfield  {author} {\bibinfo {author} {\bibfnamefont {L.}~\bibnamefont
  {Jin}}\ and\ \bibinfo {author} {\bibfnamefont {Z.}~\bibnamefont {Song}},\
  }\bibfield  {title} {\bibinfo {title} {Incident direction independent wave
  propagation and unidirectional lasing},\ }\href
  {https://doi.org/10.1103/PhysRevLett.121.073901} {\bibfield  {journal}
  {\bibinfo  {journal} {Phys. Rev. Lett.}\ }\textbf {\bibinfo {volume} {121}},\
  \bibinfo {pages} {073901} (\bibinfo {year} {2018})}\BibitemShut {NoStop}%
\bibitem [{\citenamefont {Peng}\ \emph {et~al.}(2014)\citenamefont {Peng},
  \citenamefont {\c{S} K.~{\"O}zdemir}, \citenamefont {Rotter}, \citenamefont
  {Yilmaz}, \citenamefont {Liertzer}, \citenamefont {Monifi}, \citenamefont
  {Bender}, \citenamefont {Nori},\ and\ \citenamefont {Yang}}]{peng2014}%
  \BibitemOpen
  \bibfield  {author} {\bibinfo {author} {\bibfnamefont {B.}~\bibnamefont
  {Peng}}, \bibinfo {author} {\bibnamefont {\c{S} K.~{\"O}zdemir}}, \bibinfo
  {author} {\bibfnamefont {S.}~\bibnamefont {Rotter}}, \bibinfo {author}
  {\bibfnamefont {H.}~\bibnamefont {Yilmaz}}, \bibinfo {author} {\bibfnamefont
  {M.}~\bibnamefont {Liertzer}}, \bibinfo {author} {\bibfnamefont
  {F.}~\bibnamefont {Monifi}}, \bibinfo {author} {\bibfnamefont {C.~M.}\
  \bibnamefont {Bender}}, \bibinfo {author} {\bibfnamefont {F.}~\bibnamefont
  {Nori}},\ and\ \bibinfo {author} {\bibfnamefont {L.}~\bibnamefont {Yang}},\
  }\bibfield  {title} {\bibinfo {title} {Loss-induced suppression and revival
  of lasing},\ }\href {https://doi.org/10.1126/science.1258004} {\bibfield
  {journal} {\bibinfo  {journal} {Science}\ }\textbf {\bibinfo {volume}
  {346}},\ \bibinfo {pages} {328} (\bibinfo {year} {2014})},\ \Eprint
  {https://arxiv.org/abs/https://www.science.org/doi/pdf/10.1126/science.1258004}
  {https://www.science.org/doi/pdf/10.1126/science.1258004} \BibitemShut
  {NoStop}%
\bibitem [{\citenamefont {Longhi}\ \emph {et~al.}(2015)\citenamefont {Longhi},
  \citenamefont {Gatti},\ and\ \citenamefont {Valle}}]{longhi2015}%
  \BibitemOpen
  \bibfield  {author} {\bibinfo {author} {\bibfnamefont {S.}~\bibnamefont
  {Longhi}}, \bibinfo {author} {\bibfnamefont {D.}~\bibnamefont {Gatti}},\ and\
  \bibinfo {author} {\bibfnamefont {G.~D.}\ \bibnamefont {Valle}},\ }\bibfield
  {title} {\bibinfo {title} {Robust light transport in non-{H}ermitian photonic
  lattices},\ }\href {https://doi.org/10.1038/srep13376} {\bibfield  {journal}
  {\bibinfo  {journal} {Scientific Reports}\ }\textbf {\bibinfo {volume} {5}},\
  \bibinfo {pages} {13376} (\bibinfo {year} {2015})}\BibitemShut {NoStop}%
\bibitem [{\citenamefont {Kawabata}\ \emph {et~al.}(2021)\citenamefont
  {Kawabata}, \citenamefont {Shiozaki},\ and\ \citenamefont
  {Ryu}}]{kawabata2021}%
  \BibitemOpen
  \bibfield  {author} {\bibinfo {author} {\bibfnamefont {K.}~\bibnamefont
  {Kawabata}}, \bibinfo {author} {\bibfnamefont {K.}~\bibnamefont {Shiozaki}},\
  and\ \bibinfo {author} {\bibfnamefont {S.}~\bibnamefont {Ryu}},\ }\bibfield
  {title} {\bibinfo {title} {Topological field theory of non-{H}ermitian
  systems},\ }\href {https://doi.org/10.1103/PhysRevLett.126.216405} {\bibfield
   {journal} {\bibinfo  {journal} {Phys. Rev. Lett.}\ }\textbf {\bibinfo
  {volume} {126}},\ \bibinfo {pages} {216405} (\bibinfo {year}
  {2021})}\BibitemShut {NoStop}%
\bibitem [{\citenamefont {Kasprzak}\ \emph {et~al.}(2006)\citenamefont
  {Kasprzak}, \citenamefont {Richard}, \citenamefont {Kundermann},
  \citenamefont {Baas}, \citenamefont {Jeambrun}, \citenamefont {Keeling},
  \citenamefont {Marchetti}, \citenamefont {Szyma{\'n}ska}, \citenamefont
  {Andr{\'e}}, \citenamefont {Staehli}, \citenamefont {Savona}, \citenamefont
  {Littlewood}, \citenamefont {Deveaud},\ and\ \citenamefont
  {Dang}}]{kasprzak2006}%
  \BibitemOpen
  \bibfield  {author} {\bibinfo {author} {\bibfnamefont {J.}~\bibnamefont
  {Kasprzak}}, \bibinfo {author} {\bibfnamefont {M.}~\bibnamefont {Richard}},
  \bibinfo {author} {\bibfnamefont {S.}~\bibnamefont {Kundermann}}, \bibinfo
  {author} {\bibfnamefont {A.}~\bibnamefont {Baas}}, \bibinfo {author}
  {\bibfnamefont {P.}~\bibnamefont {Jeambrun}}, \bibinfo {author}
  {\bibfnamefont {J.~M.~J.}\ \bibnamefont {Keeling}}, \bibinfo {author}
  {\bibfnamefont {F.~M.}\ \bibnamefont {Marchetti}}, \bibinfo {author}
  {\bibfnamefont {M.~H.}\ \bibnamefont {Szyma{\'n}ska}}, \bibinfo {author}
  {\bibfnamefont {R.}~\bibnamefont {Andr{\'e}}}, \bibinfo {author}
  {\bibfnamefont {J.~L.}\ \bibnamefont {Staehli}}, \bibinfo {author}
  {\bibfnamefont {V.}~\bibnamefont {Savona}}, \bibinfo {author} {\bibfnamefont
  {P.~B.}\ \bibnamefont {Littlewood}}, \bibinfo {author} {\bibfnamefont
  {B.}~\bibnamefont {Deveaud}},\ and\ \bibinfo {author} {\bibfnamefont {L.~S.}\
  \bibnamefont {Dang}},\ }\bibfield  {title} {\bibinfo {title}
  {{B}ose--{E}instein condensation of exciton polaritons},\ }\href
  {https://doi.org/10.1038/nature05131} {\bibfield  {journal} {\bibinfo
  {journal} {Nature}\ }\textbf {\bibinfo {volume} {443}},\ \bibinfo {pages}
  {409} (\bibinfo {year} {2006})}\BibitemShut {NoStop}%
\bibitem [{\citenamefont {Carusotto}\ and\ \citenamefont
  {Ciuti}(2013)}]{carusotto2013}%
  \BibitemOpen
  \bibfield  {author} {\bibinfo {author} {\bibfnamefont {I.}~\bibnamefont
  {Carusotto}}\ and\ \bibinfo {author} {\bibfnamefont {C.}~\bibnamefont
  {Ciuti}},\ }\bibfield  {title} {\bibinfo {title} {Quantum fluids of light},\
  }\href {https://doi.org/10.1103/RevModPhys.85.299} {\bibfield  {journal}
  {\bibinfo  {journal} {Rev. Mod. Phys.}\ }\textbf {\bibinfo {volume} {85}},\
  \bibinfo {pages} {299} (\bibinfo {year} {2013})}\BibitemShut {NoStop}%
\bibitem [{\citenamefont {Deng}\ \emph {et~al.}(2010)\citenamefont {Deng},
  \citenamefont {Haug},\ and\ \citenamefont {Yamamoto}}]{deng2010}%
  \BibitemOpen
  \bibfield  {author} {\bibinfo {author} {\bibfnamefont {H.}~\bibnamefont
  {Deng}}, \bibinfo {author} {\bibfnamefont {H.}~\bibnamefont {Haug}},\ and\
  \bibinfo {author} {\bibfnamefont {Y.}~\bibnamefont {Yamamoto}},\ }\bibfield
  {title} {\bibinfo {title} {Exciton-polariton {B}ose-{E}instein
  condensation},\ }\href {https://doi.org/10.1103/RevModPhys.82.1489}
  {\bibfield  {journal} {\bibinfo  {journal} {Rev. Mod. Phys.}\ }\textbf
  {\bibinfo {volume} {82}},\ \bibinfo {pages} {1489} (\bibinfo {year}
  {2010})}\BibitemShut {NoStop}%
\bibitem [{\citenamefont {Liao}\ \emph {et~al.}(2021)\citenamefont {Liao},
  \citenamefont {Leblanc}, \citenamefont {Ren}, \citenamefont {Li},
  \citenamefont {Li}, \citenamefont {Solnyshkov}, \citenamefont {Malpuech},
  \citenamefont {Yao},\ and\ \citenamefont {Fu}}]{liao2021}%
  \BibitemOpen
  \bibfield  {author} {\bibinfo {author} {\bibfnamefont {Q.}~\bibnamefont
  {Liao}}, \bibinfo {author} {\bibfnamefont {C.}~\bibnamefont {Leblanc}},
  \bibinfo {author} {\bibfnamefont {J.}~\bibnamefont {Ren}}, \bibinfo {author}
  {\bibfnamefont {F.}~\bibnamefont {Li}}, \bibinfo {author} {\bibfnamefont
  {Y.}~\bibnamefont {Li}}, \bibinfo {author} {\bibfnamefont {D.}~\bibnamefont
  {Solnyshkov}}, \bibinfo {author} {\bibfnamefont {G.}~\bibnamefont
  {Malpuech}}, \bibinfo {author} {\bibfnamefont {J.}~\bibnamefont {Yao}},\ and\
  \bibinfo {author} {\bibfnamefont {H.}~\bibnamefont {Fu}},\ }\bibfield
  {title} {\bibinfo {title} {Experimental measurement of the divergent quantum
  metric of an exceptional point},\ }\href
  {https://doi.org/10.1103/PhysRevLett.127.107402} {\bibfield  {journal}
  {\bibinfo  {journal} {Phys. Rev. Lett.}\ }\textbf {\bibinfo {volume} {127}},\
  \bibinfo {pages} {107402} (\bibinfo {year} {2021})}\BibitemShut {NoStop}%
\bibitem [{\citenamefont {Bleu}\ \emph
  {et~al.}(2018{\natexlab{a}})\citenamefont {Bleu}, \citenamefont {Malpuech},
  \citenamefont {Gao},\ and\ \citenamefont {Solnyshkov}}]{bleu2018wp}%
  \BibitemOpen
  \bibfield  {author} {\bibinfo {author} {\bibfnamefont {O.}~\bibnamefont
  {Bleu}}, \bibinfo {author} {\bibfnamefont {G.}~\bibnamefont {Malpuech}},
  \bibinfo {author} {\bibfnamefont {Y.}~\bibnamefont {Gao}},\ and\ \bibinfo
  {author} {\bibfnamefont {D.~D.}\ \bibnamefont {Solnyshkov}},\ }\bibfield
  {title} {\bibinfo {title} {Effective theory of nonadiabatic quantum evolution
  based on the quantum geometric tensor},\ }\href
  {https://doi.org/10.1103/PhysRevLett.121.020401} {\bibfield  {journal}
  {\bibinfo  {journal} {Phys. Rev. Lett.}\ }\textbf {\bibinfo {volume} {121}},\
  \bibinfo {pages} {020401} (\bibinfo {year} {2018}{\natexlab{a}})}\BibitemShut
  {NoStop}%
\bibitem [{\citenamefont {Culcer}\ \emph {et~al.}(2005)\citenamefont {Culcer},
  \citenamefont {Yao},\ and\ \citenamefont {Niu}}]{culcer2005}%
  \BibitemOpen
  \bibfield  {author} {\bibinfo {author} {\bibfnamefont {D.}~\bibnamefont
  {Culcer}}, \bibinfo {author} {\bibfnamefont {Y.}~\bibnamefont {Yao}},\ and\
  \bibinfo {author} {\bibfnamefont {Q.}~\bibnamefont {Niu}},\ }\bibfield
  {title} {\bibinfo {title} {Coherent wave-packet evolution in coupled bands},\
  }\href {https://doi.org/10.1103/PhysRevB.72.085110} {\bibfield  {journal}
  {\bibinfo  {journal} {Phys. Rev. B}\ }\textbf {\bibinfo {volume} {72}},\
  \bibinfo {pages} {085110} (\bibinfo {year} {2005})}\BibitemShut {NoStop}%
\bibitem [{\citenamefont {Leblanc}\ \emph {et~al.}(2021)\citenamefont
  {Leblanc}, \citenamefont {Malpuech},\ and\ \citenamefont
  {Solnyshkov}}]{leblanc2021}%
  \BibitemOpen
  \bibfield  {author} {\bibinfo {author} {\bibfnamefont {C.}~\bibnamefont
  {Leblanc}}, \bibinfo {author} {\bibfnamefont {G.}~\bibnamefont {Malpuech}},\
  and\ \bibinfo {author} {\bibfnamefont {D.~D.}\ \bibnamefont {Solnyshkov}},\
  }\bibfield  {title} {\bibinfo {title} {Universal semiclassical equations
  based on the quantum metric for a two-band system},\ }\href
  {https://doi.org/10.1103/PhysRevB.104.134312} {\bibfield  {journal} {\bibinfo
   {journal} {Phys. Rev. B}\ }\textbf {\bibinfo {volume} {104}},\ \bibinfo
  {pages} {134312} (\bibinfo {year} {2021})}\BibitemShut {NoStop}%
\bibitem [{\citenamefont {Solnyshkov}\ \emph {et~al.}(2021)\citenamefont
  {Solnyshkov}, \citenamefont {Leblanc}, \citenamefont {Bessonart},
  \citenamefont {Nalitov}, \citenamefont {Ren}, \citenamefont {Liao},
  \citenamefont {Li},\ and\ \citenamefont {Malpuech}}]{solnyshkov2021}%
  \BibitemOpen
  \bibfield  {author} {\bibinfo {author} {\bibfnamefont {D.~D.}\ \bibnamefont
  {Solnyshkov}}, \bibinfo {author} {\bibfnamefont {C.}~\bibnamefont {Leblanc}},
  \bibinfo {author} {\bibfnamefont {L.}~\bibnamefont {Bessonart}}, \bibinfo
  {author} {\bibfnamefont {A.}~\bibnamefont {Nalitov}}, \bibinfo {author}
  {\bibfnamefont {J.}~\bibnamefont {Ren}}, \bibinfo {author} {\bibfnamefont
  {Q.}~\bibnamefont {Liao}}, \bibinfo {author} {\bibfnamefont {F.}~\bibnamefont
  {Li}},\ and\ \bibinfo {author} {\bibfnamefont {G.}~\bibnamefont {Malpuech}},\
  }\bibfield  {title} {\bibinfo {title} {Quantum metric and wave packets at
  exceptional points in non-{H}ermitian systems},\ }\href
  {https://doi.org/10.1103/PhysRevB.103.125302} {\bibfield  {journal} {\bibinfo
   {journal} {Phys. Rev. B}\ }\textbf {\bibinfo {volume} {103}},\ \bibinfo
  {pages} {125302} (\bibinfo {year} {2021})}\BibitemShut {NoStop}%
\bibitem [{\citenamefont {Longhi}(2022)}]{longhi2022}%
  \BibitemOpen
  \bibfield  {author} {\bibinfo {author} {\bibfnamefont {S.}~\bibnamefont
  {Longhi}},\ }\bibfield  {title} {\bibinfo {title} {Non-{H}ermitian skin
  effect and self-acceleration},\ }\href
  {https://doi.org/10.1103/PhysRevB.105.245143} {\bibfield  {journal} {\bibinfo
   {journal} {Phys. Rev. B}\ }\textbf {\bibinfo {volume} {105}},\ \bibinfo
  {pages} {245143} (\bibinfo {year} {2022})}\BibitemShut {NoStop}%
\bibitem [{\citenamefont {Cilibrizzi}\ \emph {et~al.}(2016)\citenamefont
  {Cilibrizzi}, \citenamefont {Sigurdsson}, \citenamefont {Liew}, \citenamefont
  {Ohadi}, \citenamefont {Askitopoulos}, \citenamefont {Brodbeck},
  \citenamefont {Schneider}, \citenamefont {Shelykh}, \citenamefont
  {H\"ofling}, \citenamefont {Ruostekoski},\ and\ \citenamefont
  {Lagoudakis}}]{cilibrizzi2016}%
  \BibitemOpen
  \bibfield  {author} {\bibinfo {author} {\bibfnamefont {P.}~\bibnamefont
  {Cilibrizzi}}, \bibinfo {author} {\bibfnamefont {H.}~\bibnamefont
  {Sigurdsson}}, \bibinfo {author} {\bibfnamefont {T.~C.~H.}\ \bibnamefont
  {Liew}}, \bibinfo {author} {\bibfnamefont {H.}~\bibnamefont {Ohadi}},
  \bibinfo {author} {\bibfnamefont {A.}~\bibnamefont {Askitopoulos}}, \bibinfo
  {author} {\bibfnamefont {S.}~\bibnamefont {Brodbeck}}, \bibinfo {author}
  {\bibfnamefont {C.}~\bibnamefont {Schneider}}, \bibinfo {author}
  {\bibfnamefont {I.~A.}\ \bibnamefont {Shelykh}}, \bibinfo {author}
  {\bibfnamefont {S.}~\bibnamefont {H\"ofling}}, \bibinfo {author}
  {\bibfnamefont {J.}~\bibnamefont {Ruostekoski}},\ and\ \bibinfo {author}
  {\bibfnamefont {P.}~\bibnamefont {Lagoudakis}},\ }\bibfield  {title}
  {\bibinfo {title} {Half-skyrmion spin textures in polariton microcavities},\
  }\href {https://doi.org/10.1103/PhysRevB.94.045315} {\bibfield  {journal}
  {\bibinfo  {journal} {Phys. Rev. B}\ }\textbf {\bibinfo {volume} {94}},\
  \bibinfo {pages} {045315} (\bibinfo {year} {2016})}\BibitemShut {NoStop}%
\bibitem [{\citenamefont {Nagaosa}\ and\ \citenamefont
  {Tokura}(2013)}]{nagaosa2013}%
  \BibitemOpen
  \bibfield  {author} {\bibinfo {author} {\bibfnamefont {N.}~\bibnamefont
  {Nagaosa}}\ and\ \bibinfo {author} {\bibfnamefont {Y.}~\bibnamefont
  {Tokura}},\ }\bibfield  {title} {\bibinfo {title} {Topological properties and
  dynamics of magnetic skyrmions},\ }\href
  {https://doi.org/10.1038/nnano.2013.243} {\bibfield  {journal} {\bibinfo
  {journal} {Nature Nanotechnology}\ }\textbf {\bibinfo {volume} {8}},\
  \bibinfo {pages} {899} (\bibinfo {year} {2013})}\BibitemShut {NoStop}%
\bibitem [{\citenamefont {Guo}\ \emph {et~al.}(2020)\citenamefont {Guo},
  \citenamefont {Xiao}, \citenamefont {Guo}, \citenamefont {Yuan},\ and\
  \citenamefont {Fan}}]{guo2020}%
  \BibitemOpen
  \bibfield  {author} {\bibinfo {author} {\bibfnamefont {C.}~\bibnamefont
  {Guo}}, \bibinfo {author} {\bibfnamefont {M.}~\bibnamefont {Xiao}}, \bibinfo
  {author} {\bibfnamefont {Y.}~\bibnamefont {Guo}}, \bibinfo {author}
  {\bibfnamefont {L.}~\bibnamefont {Yuan}},\ and\ \bibinfo {author}
  {\bibfnamefont {S.}~\bibnamefont {Fan}},\ }\bibfield  {title} {\bibinfo
  {title} {Meron spin textures in momentum space},\ }\href
  {https://doi.org/10.1103/PhysRevLett.124.106103} {\bibfield  {journal}
  {\bibinfo  {journal} {Phys. Rev. Lett.}\ }\textbf {\bibinfo {volume} {124}},\
  \bibinfo {pages} {106103} (\bibinfo {year} {2020})}\BibitemShut {NoStop}%
\bibitem [{\citenamefont {Zhang}\ \emph {et~al.}(2021)\citenamefont {Zhang},
  \citenamefont {Xie}, \citenamefont {Du}, \citenamefont {Shi},\ and\
  \citenamefont {Yuan}}]{zhang2021}%
  \BibitemOpen
  \bibfield  {author} {\bibinfo {author} {\bibfnamefont {Q.}~\bibnamefont
  {Zhang}}, \bibinfo {author} {\bibfnamefont {Z.}~\bibnamefont {Xie}}, \bibinfo
  {author} {\bibfnamefont {L.}~\bibnamefont {Du}}, \bibinfo {author}
  {\bibfnamefont {P.}~\bibnamefont {Shi}},\ and\ \bibinfo {author}
  {\bibfnamefont {X.}~\bibnamefont {Yuan}},\ }\bibfield  {title} {\bibinfo
  {title} {Bloch-type photonic skyrmions in optical chiral multilayers},\
  }\href {https://doi.org/10.1103/PhysRevResearch.3.023109} {\bibfield
  {journal} {\bibinfo  {journal} {Phys. Rev. Research}\ }\textbf {\bibinfo
  {volume} {3}},\ \bibinfo {pages} {023109} (\bibinfo {year}
  {2021})}\BibitemShut {NoStop}%
\bibitem [{\citenamefont {G{\"o}bel}\ \emph {et~al.}(2021)\citenamefont
  {G{\"o}bel}, \citenamefont {Mertig},\ and\ \citenamefont
  {Tretiakov}}]{borge2021}%
  \BibitemOpen
  \bibfield  {author} {\bibinfo {author} {\bibfnamefont {B.}~\bibnamefont
  {G{\"o}bel}}, \bibinfo {author} {\bibfnamefont {I.}~\bibnamefont {Mertig}},\
  and\ \bibinfo {author} {\bibfnamefont {O.~A.}\ \bibnamefont {Tretiakov}},\
  }\bibfield  {title} {\bibinfo {title} {Beyond skyrmions: Review and
  perspectives of alternative magnetic quasiparticles},\ }\href
  {https://doi.org/https://doi.org/10.1016/j.physrep.2020.10.001} {\bibfield
  {journal} {\bibinfo  {journal} {Physics Reports}\ }\textbf {\bibinfo {volume}
  {895}},\ \bibinfo {pages} {1} (\bibinfo {year} {2021})},\ \bibinfo {note}
  {beyond skyrmions: Review and perspectives of alternative magnetic
  quasiparticles}\BibitemShut {NoStop}%
\bibitem [{\citenamefont {Kr\'{o}l}\ \emph {et~al.}(2021)\citenamefont
  {Kr\'{o}l}, \citenamefont {Sigurdsson}, \citenamefont {Rechci\'{n}ska},
  \citenamefont {Oliwa}, \citenamefont {Tyszka}, \citenamefont {Bardyszewski},
  \citenamefont {Opala}, \citenamefont {Matuszewski}, \citenamefont {Morawiak},
  \citenamefont {Mazur}, \citenamefont {Piecek}, \citenamefont {Kula},
  \citenamefont {Lagoudakis}, \citenamefont {Pi\c{e}tka},\ and\ \citenamefont
  {Szczytko}}]{krol2021}%
  \BibitemOpen
  \bibfield  {author} {\bibinfo {author} {\bibfnamefont {M.}~\bibnamefont
  {Kr\'{o}l}}, \bibinfo {author} {\bibfnamefont {H.}~\bibnamefont
  {Sigurdsson}}, \bibinfo {author} {\bibfnamefont {K.}~\bibnamefont
  {Rechci\'{n}ska}}, \bibinfo {author} {\bibfnamefont {P.}~\bibnamefont
  {Oliwa}}, \bibinfo {author} {\bibfnamefont {K.}~\bibnamefont {Tyszka}},
  \bibinfo {author} {\bibfnamefont {W.}~\bibnamefont {Bardyszewski}}, \bibinfo
  {author} {\bibfnamefont {A.}~\bibnamefont {Opala}}, \bibinfo {author}
  {\bibfnamefont {M.}~\bibnamefont {Matuszewski}}, \bibinfo {author}
  {\bibfnamefont {P.}~\bibnamefont {Morawiak}}, \bibinfo {author}
  {\bibfnamefont {R.}~\bibnamefont {Mazur}}, \bibinfo {author} {\bibfnamefont
  {W.}~\bibnamefont {Piecek}}, \bibinfo {author} {\bibfnamefont
  {P.}~\bibnamefont {Kula}}, \bibinfo {author} {\bibfnamefont {P.~G.}\
  \bibnamefont {Lagoudakis}}, \bibinfo {author} {\bibfnamefont
  {B.}~\bibnamefont {Pi\c{e}tka}},\ and\ \bibinfo {author} {\bibfnamefont
  {J.}~\bibnamefont {Szczytko}},\ }\bibfield  {title} {\bibinfo {title}
  {Observation of second-order meron polarization textures in optical
  microcavities},\ }\href {https://doi.org/10.1364/OPTICA.414891} {\bibfield
  {journal} {\bibinfo  {journal} {Optica}\ }\textbf {\bibinfo {volume} {8}},\
  \bibinfo {pages} {255} (\bibinfo {year} {2021})}\BibitemShut {NoStop}%
\bibitem [{\citenamefont {Flayac}\ \emph {et~al.}(2013)\citenamefont {Flayac},
  \citenamefont {Solnyshkov}, \citenamefont {Shelykh},\ and\ \citenamefont
  {Malpuech}}]{flayac2013}%
  \BibitemOpen
  \bibfield  {author} {\bibinfo {author} {\bibfnamefont {H.}~\bibnamefont
  {Flayac}}, \bibinfo {author} {\bibfnamefont {D.~D.}\ \bibnamefont
  {Solnyshkov}}, \bibinfo {author} {\bibfnamefont {I.~A.}\ \bibnamefont
  {Shelykh}},\ and\ \bibinfo {author} {\bibfnamefont {G.}~\bibnamefont
  {Malpuech}},\ }\bibfield  {title} {\bibinfo {title} {Transmutation of
  skyrmions to half-solitons driven by the nonlinear optical spin {H}all
  effect},\ }\href {https://doi.org/10.1103/PhysRevLett.110.016404} {\bibfield
  {journal} {\bibinfo  {journal} {Phys. Rev. Lett.}\ }\textbf {\bibinfo
  {volume} {110}},\ \bibinfo {pages} {016404} (\bibinfo {year}
  {2013})}\BibitemShut {NoStop}%
\bibitem [{\citenamefont {Vishnevsky}\ \emph {et~al.}(2013)\citenamefont
  {Vishnevsky}, \citenamefont {Flayac}, \citenamefont {Nalitov}, \citenamefont
  {Solnyshkov}, \citenamefont {Gippius},\ and\ \citenamefont
  {Malpuech}}]{vishnevsky2013}%
  \BibitemOpen
  \bibfield  {author} {\bibinfo {author} {\bibfnamefont {D.~V.}\ \bibnamefont
  {Vishnevsky}}, \bibinfo {author} {\bibfnamefont {H.}~\bibnamefont {Flayac}},
  \bibinfo {author} {\bibfnamefont {A.~V.}\ \bibnamefont {Nalitov}}, \bibinfo
  {author} {\bibfnamefont {D.~D.}\ \bibnamefont {Solnyshkov}}, \bibinfo
  {author} {\bibfnamefont {N.~A.}\ \bibnamefont {Gippius}},\ and\ \bibinfo
  {author} {\bibfnamefont {G.}~\bibnamefont {Malpuech}},\ }\bibfield  {title}
  {\bibinfo {title} {Skyrmion formation and optical spin-{H}all effect in an
  expanding coherent cloud of indirect excitons},\ }\href
  {https://doi.org/10.1103/PhysRevLett.110.246404} {\bibfield  {journal}
  {\bibinfo  {journal} {Phys. Rev. Lett.}\ }\textbf {\bibinfo {volume} {110}},\
  \bibinfo {pages} {246404} (\bibinfo {year} {2013})}\BibitemShut {NoStop}%
\bibitem [{\citenamefont {Mohanta}\ \emph {et~al.}(2017)\citenamefont
  {Mohanta}, \citenamefont {Kampf},\ and\ \citenamefont {Kopp}}]{mohanta2017}%
  \BibitemOpen
  \bibfield  {author} {\bibinfo {author} {\bibfnamefont {N.}~\bibnamefont
  {Mohanta}}, \bibinfo {author} {\bibfnamefont {A.~P.}\ \bibnamefont {Kampf}},\
  and\ \bibinfo {author} {\bibfnamefont {T.}~\bibnamefont {Kopp}},\ }\bibfield
  {title} {\bibinfo {title} {Emergent momentum-space skyrmion texture on the
  surface of topological insulators},\ }\href
  {https://doi.org/10.1038/srep45664} {\bibfield  {journal} {\bibinfo
  {journal} {Scientific Reports}\ }\textbf {\bibinfo {volume} {7}},\ \bibinfo
  {pages} {45664} (\bibinfo {year} {2017})}\BibitemShut {NoStop}%
\bibitem [{\citenamefont {Ter\ifmmode~\mbox{\c{c}}\else \c{c}\fi{}as}\ \emph
  {et~al.}(2014)\citenamefont {Ter\ifmmode~\mbox{\c{c}}\else \c{c}\fi{}as},
  \citenamefont {Flayac}, \citenamefont {Solnyshkov},\ and\ \citenamefont
  {Malpuech}}]{tercas2014}%
  \BibitemOpen
  \bibfield  {author} {\bibinfo {author} {\bibfnamefont {H.}~\bibnamefont
  {Ter\ifmmode~\mbox{\c{c}}\else \c{c}\fi{}as}}, \bibinfo {author}
  {\bibfnamefont {H.}~\bibnamefont {Flayac}}, \bibinfo {author} {\bibfnamefont
  {D.~D.}\ \bibnamefont {Solnyshkov}},\ and\ \bibinfo {author} {\bibfnamefont
  {G.}~\bibnamefont {Malpuech}},\ }\bibfield  {title} {\bibinfo {title}
  {Non-{A}belian gauge fields in photonic cavities and photonic superfluids},\
  }\href {https://doi.org/10.1103/PhysRevLett.112.066402} {\bibfield  {journal}
  {\bibinfo  {journal} {Phys. Rev. Lett.}\ }\textbf {\bibinfo {volume} {112}},\
  \bibinfo {pages} {066402} (\bibinfo {year} {2014})}\BibitemShut {NoStop}%
\bibitem [{\citenamefont {Klembt}\ \emph {et~al.}(2018)\citenamefont {Klembt},
  \citenamefont {Harder}, \citenamefont {Egorov}, \citenamefont {Winkler},
  \citenamefont {Ge}, \citenamefont {Bandres}, \citenamefont {Emmerling},
  \citenamefont {Worschech}, \citenamefont {Liew}, \citenamefont {Segev},
  \citenamefont {Schneider},\ and\ \citenamefont {H{\"o}fling}}]{klembt2018}%
  \BibitemOpen
  \bibfield  {author} {\bibinfo {author} {\bibfnamefont {S.}~\bibnamefont
  {Klembt}}, \bibinfo {author} {\bibfnamefont {T.~H.}\ \bibnamefont {Harder}},
  \bibinfo {author} {\bibfnamefont {O.~A.}\ \bibnamefont {Egorov}}, \bibinfo
  {author} {\bibfnamefont {K.}~\bibnamefont {Winkler}}, \bibinfo {author}
  {\bibfnamefont {R.}~\bibnamefont {Ge}}, \bibinfo {author} {\bibfnamefont
  {M.~A.}\ \bibnamefont {Bandres}}, \bibinfo {author} {\bibfnamefont
  {M.}~\bibnamefont {Emmerling}}, \bibinfo {author} {\bibfnamefont
  {L.}~\bibnamefont {Worschech}}, \bibinfo {author} {\bibfnamefont {T.~C.~H.}\
  \bibnamefont {Liew}}, \bibinfo {author} {\bibfnamefont {M.}~\bibnamefont
  {Segev}}, \bibinfo {author} {\bibfnamefont {C.}~\bibnamefont {Schneider}},\
  and\ \bibinfo {author} {\bibfnamefont {S.}~\bibnamefont {H{\"o}fling}},\
  }\bibfield  {title} {\bibinfo {title} {Exciton-polariton topological
  insulator},\ }\href {https://doi.org/10.1038/s41586-018-0601-5} {\bibfield
  {journal} {\bibinfo  {journal} {Nature}\ }\textbf {\bibinfo {volume} {562}},\
  \bibinfo {pages} {552} (\bibinfo {year} {2018})}\BibitemShut {NoStop}%
\bibitem [{\citenamefont {Zhou}\ \emph {et~al.}(2018)\citenamefont {Zhou},
  \citenamefont {Peng}, \citenamefont {Yoon}, \citenamefont {Hsu},
  \citenamefont {Nelson}, \citenamefont {Fu}, \citenamefont {Joannopoulos},
  \citenamefont {Soljačić},\ and\ \citenamefont {Zhen}}]{zhou2018}%
  \BibitemOpen
  \bibfield  {author} {\bibinfo {author} {\bibfnamefont {H.}~\bibnamefont
  {Zhou}}, \bibinfo {author} {\bibfnamefont {C.}~\bibnamefont {Peng}}, \bibinfo
  {author} {\bibfnamefont {Y.}~\bibnamefont {Yoon}}, \bibinfo {author}
  {\bibfnamefont {C.~W.}\ \bibnamefont {Hsu}}, \bibinfo {author} {\bibfnamefont
  {K.~A.}\ \bibnamefont {Nelson}}, \bibinfo {author} {\bibfnamefont
  {L.}~\bibnamefont {Fu}}, \bibinfo {author} {\bibfnamefont {J.~D.}\
  \bibnamefont {Joannopoulos}}, \bibinfo {author} {\bibfnamefont
  {M.}~\bibnamefont {Soljačić}},\ and\ \bibinfo {author} {\bibfnamefont
  {B.}~\bibnamefont {Zhen}},\ }\bibfield  {title} {\bibinfo {title}
  {Observation of bulk {F}ermi arc and polarization half charge from paired
  exceptional points},\ }\href {https://doi.org/10.1126/science.aap9859}
  {\bibfield  {journal} {\bibinfo  {journal} {Science}\ }\textbf {\bibinfo
  {volume} {359}},\ \bibinfo {pages} {1009} (\bibinfo {year} {2018})},\ \Eprint
  {https://arxiv.org/abs/https://www.science.org/doi/pdf/10.1126/science.aap9859}
  {https://www.science.org/doi/pdf/10.1126/science.aap9859} \BibitemShut
  {NoStop}%
\bibitem [{\citenamefont {Donati}\ \emph {et~al.}(2016)\citenamefont {Donati},
  \citenamefont {Dominici}, \citenamefont {Dagvadorj}, \citenamefont
  {Ballarini}, \citenamefont {Giorgi}, \citenamefont {Bramati}, \citenamefont
  {Gigli}, \citenamefont {Rubo}, \citenamefont {Szyma{\'n}ska},\ and\
  \citenamefont {Sanvitto}}]{stefano2016}%
  \BibitemOpen
  \bibfield  {author} {\bibinfo {author} {\bibfnamefont {S.}~\bibnamefont
  {Donati}}, \bibinfo {author} {\bibfnamefont {L.}~\bibnamefont {Dominici}},
  \bibinfo {author} {\bibfnamefont {G.}~\bibnamefont {Dagvadorj}}, \bibinfo
  {author} {\bibfnamefont {D.}~\bibnamefont {Ballarini}}, \bibinfo {author}
  {\bibfnamefont {M.~D.}\ \bibnamefont {Giorgi}}, \bibinfo {author}
  {\bibfnamefont {A.}~\bibnamefont {Bramati}}, \bibinfo {author} {\bibfnamefont
  {G.}~\bibnamefont {Gigli}}, \bibinfo {author} {\bibfnamefont {Y.~G.}\
  \bibnamefont {Rubo}}, \bibinfo {author} {\bibfnamefont {M.~H.}\ \bibnamefont
  {Szyma{\'n}ska}},\ and\ \bibinfo {author} {\bibfnamefont {D.}~\bibnamefont
  {Sanvitto}},\ }\bibfield  {title} {\bibinfo {title} {Twist of generalized
  skyrmions and spin vortices in a polariton superfluid},\ }\href
  {https://doi.org/10.1073/pnas.1610123114} {\bibfield  {journal} {\bibinfo
  {journal} {Proceedings of the National Academy of Sciences}\ }\textbf
  {\bibinfo {volume} {113}},\ \bibinfo {pages} {14926} (\bibinfo {year}
  {2016})},\ \Eprint
  {https://arxiv.org/abs/https://www.pnas.org/doi/pdf/10.1073/pnas.1610123114}
  {https://www.pnas.org/doi/pdf/10.1073/pnas.1610123114} \BibitemShut {NoStop}%
\bibitem [{\citenamefont {Everschor-Sitte}\ \emph {et~al.}(2018)\citenamefont
  {Everschor-Sitte}, \citenamefont {Masell}, \citenamefont {Reeve},\ and\
  \citenamefont {Kl{\"a}ui}}]{everschor-sitte2018}%
  \BibitemOpen
  \bibfield  {author} {\bibinfo {author} {\bibfnamefont {K.}~\bibnamefont
  {Everschor-Sitte}}, \bibinfo {author} {\bibfnamefont {J.}~\bibnamefont
  {Masell}}, \bibinfo {author} {\bibfnamefont {R.~M.}\ \bibnamefont {Reeve}},\
  and\ \bibinfo {author} {\bibfnamefont {M.}~\bibnamefont {Kl{\"a}ui}},\
  }\bibfield  {title} {\bibinfo {title} {Perspective: Magnetic
  skyrmions---overview of recent progress in an active research field},\ }\href
  {https://doi.org/10.1063/1.5048972} {\bibfield  {journal} {\bibinfo
  {journal} {Journal of Applied Physics}\ }\textbf {\bibinfo {volume} {124}},\
  \bibinfo {pages} {240901} (\bibinfo {year} {2018})},\ \Eprint
  {https://arxiv.org/abs/https://doi.org/10.1063/1.5048972}
  {https://doi.org/10.1063/1.5048972} \BibitemShut {NoStop}%
\bibitem [{\citenamefont {Kovalev}\ and\ \citenamefont
  {Sandhoefner}(2018)}]{kovalev2018}%
  \BibitemOpen
  \bibfield  {author} {\bibinfo {author} {\bibfnamefont {A.~A.}\ \bibnamefont
  {Kovalev}}\ and\ \bibinfo {author} {\bibfnamefont {S.}~\bibnamefont
  {Sandhoefner}},\ }\bibfield  {title} {\bibinfo {title} {Skyrmions and
  antiskyrmions in quasi-two-dimensional magnets},\ }\bibfield  {journal}
  {\bibinfo  {journal} {Frontiers in Physics}\ }\textbf {\bibinfo {volume}
  {6}},\ \href {https://doi.org/10.3389/fphy.2018.00098}
  {10.3389/fphy.2018.00098} (\bibinfo {year} {2018})\BibitemShut {NoStop}%
\bibitem [{\citenamefont {Tretiakov}\ and\ \citenamefont
  {Tchernyshyov}(2007)}]{tretiakov2007}%
  \BibitemOpen
  \bibfield  {author} {\bibinfo {author} {\bibfnamefont {O.~A.}\ \bibnamefont
  {Tretiakov}}\ and\ \bibinfo {author} {\bibfnamefont {O.}~\bibnamefont
  {Tchernyshyov}},\ }\bibfield  {title} {\bibinfo {title} {Vortices in thin
  ferromagnetic films and the skyrmion number},\ }\href
  {https://doi.org/10.1103/PhysRevB.75.012408} {\bibfield  {journal} {\bibinfo
  {journal} {Phys. Rev. B}\ }\textbf {\bibinfo {volume} {75}},\ \bibinfo
  {pages} {012408} (\bibinfo {year} {2007})}\BibitemShut {NoStop}%
\bibitem [{\citenamefont {Hertel}\ and\ \citenamefont
  {Schneider}(2006)}]{hertel2006}%
  \BibitemOpen
  \bibfield  {author} {\bibinfo {author} {\bibfnamefont {R.}~\bibnamefont
  {Hertel}}\ and\ \bibinfo {author} {\bibfnamefont {C.~M.}\ \bibnamefont
  {Schneider}},\ }\bibfield  {title} {\bibinfo {title} {Exchange explosions:
  Magnetization dynamics during vortex-antivortex annihilation},\ }\href
  {https://doi.org/10.1103/PhysRevLett.97.177202} {\bibfield  {journal}
  {\bibinfo  {journal} {Phys. Rev. Lett.}\ }\textbf {\bibinfo {volume} {97}},\
  \bibinfo {pages} {177202} (\bibinfo {year} {2006})}\BibitemShut {NoStop}%
\bibitem [{\citenamefont {Richter}\ \emph {et~al.}(2019)\citenamefont
  {Richter}, \citenamefont {Zirnstein}, \citenamefont {Z\'u\~niga P\'erez},
  \citenamefont {Kr\"uger}, \citenamefont {Deparis}, \citenamefont {Trefflich},
  \citenamefont {Sturm}, \citenamefont {Rosenow}, \citenamefont {Grundmann},\
  and\ \citenamefont {Schmidt-Grund}}]{richter2019}%
  \BibitemOpen
  \bibfield  {author} {\bibinfo {author} {\bibfnamefont {S.}~\bibnamefont
  {Richter}}, \bibinfo {author} {\bibfnamefont {H.-G.}\ \bibnamefont
  {Zirnstein}}, \bibinfo {author} {\bibfnamefont {J.}~\bibnamefont {Z\'u\~niga
  P\'erez}}, \bibinfo {author} {\bibfnamefont {E.}~\bibnamefont {Kr\"uger}},
  \bibinfo {author} {\bibfnamefont {C.}~\bibnamefont {Deparis}}, \bibinfo
  {author} {\bibfnamefont {L.}~\bibnamefont {Trefflich}}, \bibinfo {author}
  {\bibfnamefont {C.}~\bibnamefont {Sturm}}, \bibinfo {author} {\bibfnamefont
  {B.}~\bibnamefont {Rosenow}}, \bibinfo {author} {\bibfnamefont
  {M.}~\bibnamefont {Grundmann}},\ and\ \bibinfo {author} {\bibfnamefont
  {R.}~\bibnamefont {Schmidt-Grund}},\ }\bibfield  {title} {\bibinfo {title}
  {Voigt exceptional points in an anisotropic {ZnO}-based planar microcavity:
  Square-root topology, polarization vortices, and circularity},\ }\href
  {https://doi.org/10.1103/PhysRevLett.123.227401} {\bibfield  {journal}
  {\bibinfo  {journal} {Phys. Rev. Lett.}\ }\textbf {\bibinfo {volume} {123}},\
  \bibinfo {pages} {227401} (\bibinfo {year} {2019})}\BibitemShut {NoStop}%
\bibitem [{\citenamefont {Kr{\'o}l}\ \emph {et~al.}(2022)\citenamefont
  {Kr{\'o}l}, \citenamefont {Septembre}, \citenamefont {Oliwa}, \citenamefont
  {K\c{e}dziora}, \citenamefont {{\L}empicka-Mirek}, \citenamefont
  {Muszy{\'n}ski}, \citenamefont {Mazur}, \citenamefont {Morawiak},
  \citenamefont {Piecek}, \citenamefont {Kula}, \citenamefont {Bardyszewski},
  \citenamefont {Lagoudakis}, \citenamefont {Solnyshkov}, \citenamefont
  {Malpuech}, \citenamefont {Pi\c{e}tka},\ and\ \citenamefont
  {Szczytko}}]{krol2022}%
  \BibitemOpen
  \bibfield  {author} {\bibinfo {author} {\bibfnamefont {M.}~\bibnamefont
  {Kr{\'o}l}}, \bibinfo {author} {\bibfnamefont {I.}~\bibnamefont {Septembre}},
  \bibinfo {author} {\bibfnamefont {P.}~\bibnamefont {Oliwa}}, \bibinfo
  {author} {\bibfnamefont {M.}~\bibnamefont {K\c{e}dziora}}, \bibinfo {author}
  {\bibfnamefont {K.}~\bibnamefont {{\L}empicka-Mirek}}, \bibinfo {author}
  {\bibfnamefont {M.}~\bibnamefont {Muszy{\'n}ski}}, \bibinfo {author}
  {\bibfnamefont {R.}~\bibnamefont {Mazur}}, \bibinfo {author} {\bibfnamefont
  {P.}~\bibnamefont {Morawiak}}, \bibinfo {author} {\bibfnamefont
  {W.}~\bibnamefont {Piecek}}, \bibinfo {author} {\bibfnamefont
  {P.}~\bibnamefont {Kula}}, \bibinfo {author} {\bibfnamefont {W.}~\bibnamefont
  {Bardyszewski}}, \bibinfo {author} {\bibfnamefont {P.~G.}\ \bibnamefont
  {Lagoudakis}}, \bibinfo {author} {\bibfnamefont {D.~D.}\ \bibnamefont
  {Solnyshkov}}, \bibinfo {author} {\bibfnamefont {G.}~\bibnamefont
  {Malpuech}}, \bibinfo {author} {\bibfnamefont {B.}~\bibnamefont
  {Pi\c{e}tka}},\ and\ \bibinfo {author} {\bibfnamefont {J.}~\bibnamefont
  {Szczytko}},\ }\bibfield  {title} {\bibinfo {title} {Annihilation of
  exceptional points from different {D}irac valleys in a 2{D} photonic
  system},\ }\href {https://doi.org/10.1038/s41467-022-33001-9} {\bibfield
  {journal} {\bibinfo  {journal} {Nature Communications}\ }\textbf {\bibinfo
  {volume} {13}},\ \bibinfo {pages} {5340} (\bibinfo {year}
  {2022})}\BibitemShut {NoStop}%
\bibitem [{\citenamefont {Bleu}\ \emph
  {et~al.}(2018{\natexlab{b}})\citenamefont {Bleu}, \citenamefont
  {Solnyshkov},\ and\ \citenamefont {Malpuech}}]{bleu2018}%
  \BibitemOpen
  \bibfield  {author} {\bibinfo {author} {\bibfnamefont {O.}~\bibnamefont
  {Bleu}}, \bibinfo {author} {\bibfnamefont {D.~D.}\ \bibnamefont
  {Solnyshkov}},\ and\ \bibinfo {author} {\bibfnamefont {G.}~\bibnamefont
  {Malpuech}},\ }\bibfield  {title} {\bibinfo {title} {Measuring the quantum
  geometric tensor in two-dimensional photonic and exciton-polariton systems},\
  }\href {https://doi.org/10.1103/PhysRevB.97.195422} {\bibfield  {journal}
  {\bibinfo  {journal} {Phys. Rev. B}\ }\textbf {\bibinfo {volume} {97}},\
  \bibinfo {pages} {195422} (\bibinfo {year} {2018}{\natexlab{b}})}\BibitemShut
  {NoStop}%
\bibitem [{\citenamefont {Novoselov}\ \emph {et~al.}(2004)\citenamefont
  {Novoselov}, \citenamefont {Geim}, \citenamefont {Morozov}, \citenamefont
  {Jiang}, \citenamefont {Zhang}, \citenamefont {Dubonos}, \citenamefont
  {Grigorieva},\ and\ \citenamefont {Firsov}}]{novoselov2004}%
  \BibitemOpen
  \bibfield  {author} {\bibinfo {author} {\bibfnamefont {K.~S.}\ \bibnamefont
  {Novoselov}}, \bibinfo {author} {\bibfnamefont {A.~K.}\ \bibnamefont {Geim}},
  \bibinfo {author} {\bibfnamefont {S.~V.}\ \bibnamefont {Morozov}}, \bibinfo
  {author} {\bibfnamefont {D.}~\bibnamefont {Jiang}}, \bibinfo {author}
  {\bibfnamefont {Y.}~\bibnamefont {Zhang}}, \bibinfo {author} {\bibfnamefont
  {S.~V.}\ \bibnamefont {Dubonos}}, \bibinfo {author} {\bibfnamefont {I.~V.}\
  \bibnamefont {Grigorieva}},\ and\ \bibinfo {author} {\bibfnamefont {A.~A.}\
  \bibnamefont {Firsov}},\ }\bibfield  {title} {\bibinfo {title} {Electric
  field effect in atomically thin carbon films},\ }\href
  {https://doi.org/10.1126/science.1102896} {\bibfield  {journal} {\bibinfo
  {journal} {Science}\ }\textbf {\bibinfo {volume} {306}},\ \bibinfo {pages}
  {666} (\bibinfo {year} {2004})},\ \Eprint
  {https://arxiv.org/abs/https://www.science.org/doi/pdf/10.1126/science.1102896}
  {https://www.science.org/doi/pdf/10.1126/science.1102896} \BibitemShut
  {NoStop}%
\bibitem [{\citenamefont {Novoselov}\ \emph {et~al.}(2005)\citenamefont
  {Novoselov}, \citenamefont {Geim}, \citenamefont {Morozov}, \citenamefont
  {Jiang}, \citenamefont {Katsnelson}, \citenamefont {Grigorieva},
  \citenamefont {Dubonos},\ and\ \citenamefont {Firsov}}]{novoselov2005}%
  \BibitemOpen
  \bibfield  {author} {\bibinfo {author} {\bibfnamefont {K.~S.}\ \bibnamefont
  {Novoselov}}, \bibinfo {author} {\bibfnamefont {A.~K.}\ \bibnamefont {Geim}},
  \bibinfo {author} {\bibfnamefont {S.~V.}\ \bibnamefont {Morozov}}, \bibinfo
  {author} {\bibfnamefont {D.}~\bibnamefont {Jiang}}, \bibinfo {author}
  {\bibfnamefont {M.~I.}\ \bibnamefont {Katsnelson}}, \bibinfo {author}
  {\bibfnamefont {I.~V.}\ \bibnamefont {Grigorieva}}, \bibinfo {author}
  {\bibfnamefont {S.~V.}\ \bibnamefont {Dubonos}},\ and\ \bibinfo {author}
  {\bibfnamefont {A.~A.}\ \bibnamefont {Firsov}},\ }\bibfield  {title}
  {\bibinfo {title} {Two-dimensional gas of massless {D}irac fermions in
  graphene},\ }\href {https://doi.org/10.1038/nature04233} {\bibfield
  {journal} {\bibinfo  {journal} {Nature}\ }\textbf {\bibinfo {volume} {438}},\
  \bibinfo {pages} {197} (\bibinfo {year} {2005})}\BibitemShut {NoStop}%
\bibitem [{\citenamefont {Zhang}\ \emph {et~al.}(2005)\citenamefont {Zhang},
  \citenamefont {Tan}, \citenamefont {Stormer},\ and\ \citenamefont
  {Kim}}]{zhang2005}%
  \BibitemOpen
  \bibfield  {author} {\bibinfo {author} {\bibfnamefont {Y.}~\bibnamefont
  {Zhang}}, \bibinfo {author} {\bibfnamefont {Y.-W.}\ \bibnamefont {Tan}},
  \bibinfo {author} {\bibfnamefont {H.~L.}\ \bibnamefont {Stormer}},\ and\
  \bibinfo {author} {\bibfnamefont {P.}~\bibnamefont {Kim}},\ }\bibfield
  {title} {\bibinfo {title} {Experimental observation of the quantum {H}all
  effect and {B}erry's phase in graphene},\ }\href
  {https://doi.org/10.1038/nature04235} {\bibfield  {journal} {\bibinfo
  {journal} {Nature}\ }\textbf {\bibinfo {volume} {438}},\ \bibinfo {pages}
  {201} (\bibinfo {year} {2005})}\BibitemShut {NoStop}%
\bibitem [{\citenamefont {Castro~Neto}\ \emph {et~al.}(2009)\citenamefont
  {Castro~Neto}, \citenamefont {Guinea}, \citenamefont {Peres}, \citenamefont
  {Novoselov},\ and\ \citenamefont {Geim}}]{castro2009}%
  \BibitemOpen
  \bibfield  {author} {\bibinfo {author} {\bibfnamefont {A.~H.}\ \bibnamefont
  {Castro~Neto}}, \bibinfo {author} {\bibfnamefont {F.}~\bibnamefont {Guinea}},
  \bibinfo {author} {\bibfnamefont {N.~M.~R.}\ \bibnamefont {Peres}}, \bibinfo
  {author} {\bibfnamefont {K.~S.}\ \bibnamefont {Novoselov}},\ and\ \bibinfo
  {author} {\bibfnamefont {A.~K.}\ \bibnamefont {Geim}},\ }\bibfield  {title}
  {\bibinfo {title} {The electronic properties of graphene},\ }\href
  {https://doi.org/10.1103/RevModPhys.81.109} {\bibfield  {journal} {\bibinfo
  {journal} {Rev. Mod. Phys.}\ }\textbf {\bibinfo {volume} {81}},\ \bibinfo
  {pages} {109} (\bibinfo {year} {2009})}\BibitemShut {NoStop}%
\bibitem [{\citenamefont {Armitage}\ \emph {et~al.}(2018)\citenamefont
  {Armitage}, \citenamefont {Mele},\ and\ \citenamefont
  {Vishwanath}}]{armitage2018}%
  \BibitemOpen
  \bibfield  {author} {\bibinfo {author} {\bibfnamefont {N.~P.}\ \bibnamefont
  {Armitage}}, \bibinfo {author} {\bibfnamefont {E.~J.}\ \bibnamefont {Mele}},\
  and\ \bibinfo {author} {\bibfnamefont {A.}~\bibnamefont {Vishwanath}},\
  }\bibfield  {title} {\bibinfo {title} {Weyl and {D}irac semimetals in
  three-dimensional solids},\ }\href
  {https://doi.org/10.1103/RevModPhys.90.015001} {\bibfield  {journal}
  {\bibinfo  {journal} {Rev. Mod. Phys.}\ }\textbf {\bibinfo {volume} {90}},\
  \bibinfo {pages} {015001} (\bibinfo {year} {2018})}\BibitemShut {NoStop}%
\bibitem [{\citenamefont {Hamilton}(1837)}]{hamilton1837}%
  \BibitemOpen
  \bibfield  {author} {\bibinfo {author} {\bibfnamefont {W.~R.}\ \bibnamefont
  {Hamilton}},\ }\bibfield  {title} {\bibinfo {title} {Third supplement to an
  essay on systems of rays},\ }\href@noop {} {\bibfield  {journal} {\bibinfo
  {journal} {Transactions of the Royal Irish Academy}\ }\textbf {\bibinfo
  {volume} {17}},\ \bibinfo {pages} {1} (\bibinfo {year} {1837})}\BibitemShut
  {NoStop}%
\bibitem [{\citenamefont {Lloyd}(1837)}]{lloyd1837}%
  \BibitemOpen
  \bibfield  {author} {\bibinfo {author} {\bibfnamefont {H.}~\bibnamefont
  {Lloyd}},\ }\bibfield  {title} {\bibinfo {title} {On the phenomena presented
  by light in its passage along the axes of biaxial crystals},\ }\href@noop {}
  {\bibfield  {journal} {\bibinfo  {journal} {Transactions of the Royal Irish
  Academy}\ }\textbf {\bibinfo {volume} {17}},\ \bibinfo {pages} {145}
  (\bibinfo {year} {1837})}\BibitemShut {NoStop}%
\bibitem [{\citenamefont {Kawabata}\ \emph {et~al.}(2018)\citenamefont
  {Kawabata}, \citenamefont {Shiozaki},\ and\ \citenamefont
  {Ueda}}]{kawabata2018}%
  \BibitemOpen
  \bibfield  {author} {\bibinfo {author} {\bibfnamefont {K.}~\bibnamefont
  {Kawabata}}, \bibinfo {author} {\bibfnamefont {K.}~\bibnamefont {Shiozaki}},\
  and\ \bibinfo {author} {\bibfnamefont {M.}~\bibnamefont {Ueda}},\ }\bibfield
  {title} {\bibinfo {title} {Anomalous helical edge states in a non-{H}ermitian
  {C}hern insulator},\ }\href {https://doi.org/10.1103/PhysRevB.98.165148}
  {\bibfield  {journal} {\bibinfo  {journal} {Phys. Rev. B}\ }\textbf {\bibinfo
  {volume} {98}},\ \bibinfo {pages} {165148} (\bibinfo {year}
  {2018})}\BibitemShut {NoStop}%
\end{thebibliography}%

\end{document}